\RequirePackage{lineno}
 \documentclass[onecolumn,superscriptaddress,aps,pre]{revtex4}
\usepackage{color}
\usepackage{amsmath}
\usepackage{amssymb}
\usepackage{graphicx}
\usepackage{subfigure}
\usepackage{natbib}
\usepackage{comment}
\usepackage{cancel}

\newcommand{\now}{\text{now}}
\newcommand{\notnow}{\cancel{\text{now}}}
\newcommand{\new}{\text{new}}
\newcommand{\old}{\text{old}}
\newcommand{\ess}{\text{ess}}

\newcommand{\SI}{Supplementary Information}

\begin{document}

\title{Evolutionary learning of microbial populations in partially predictable environments}

\author{Roaa Mohmmed Yagb Omer}
\affiliation{International  School  for  Advanced  Studies  (SISSA),  Via Bonomea 265, 34136 Trieste, Italy}

\author{Onofrio Mazzarisi}
\affiliation{The National Institute of Oceanography and Applied Geophysics (OGS), Borgo Grotta Gigante, 42/c, 34010 Sgonico TS, Italy}
\affiliation{The Abdus Salam International Centre for Theoretical Physics (ICTP), Strada Costiera 11, 34014 Trieste, Italy}

\author{Martina Dal Bello}
\affiliation{Department of Ecology and Evolutionary Biology, and Microbial Sciences Institute, Yale University, New Haven, CT 06511, USA.}

\author{Jacopo Grilli}
\affiliation{The Abdus Salam International Centre for Theoretical Physics (ICTP), Strada Costiera 11, 34014 Trieste, Italy}
\email{jgrilli@ictp.it}

\begin{abstract}
Populations evolving in fluctuating environments face the fundamental challenge of balancing adaptation to current conditions against preparation for uncertain futures. Here, we study microbial evolution in partially predictable environments using proteome allocation models that capture the trade-off between growth rate and lag time during environmental transitions. We demonstrate that evolution drives populations toward an evolutionary stable allocation strategy that minimizes resource depletion time, thereby balancing faster growth with shorter adaptation delays. In environments with temporal structure, populations evolve to learn the statistical patterns of environmental transitions through proteome pre-allocation, with the evolved allocations reflecting the transition probabilities between conditions. Our framework reveals how microbial populations can extract and exploit environmental predictability without explicit neural computation, using the proteome as a distributed memory system that encodes environmental patterns. This work demonstrates how information-theoretic principles govern cellular resource allocation and provides a mechanistic foundation for understanding learning-like behavior in evolving biological systems.
\end{abstract}

\maketitle

\section{Introduction}
Populations, from microbes to macro-organisms,  experience changing environmental conditions at multiple temporal and spatial scales.
Nutrients are not constantly available, and their concentrations can vary widely~\cite{moore2013processes}. The chemical and physical conditions of the environment change over multiple spatial and temporal scales, due to both extrinsic environmental drivers and the interactions with other organisms. For instance, in the ocean, microbes experience nutrient level fluctuations both at the scale of a few micrometers~\cite{stocker2015100} --- driven by turbulence, motility, consumption and cross-feeding ---
and at the scale of tens of kilometers, during blooms~\cite{kuhlisch2024algal}. The presence and abundance of other organisms, with which a population ecologically interacts, are also variable. These factors influence the ability of an individual to survive and reproduce, affecting fitness. Natural selection operates in fluctuating, uncertain environments, with organisms trying adapt to these variable conditions.

In contrast, textbook population genetics focuses on constant, or slowly varying, conditions~\cite{hartl1997principles}, where population growth and abundance is independent of the structure if its genetic variation. In the case of microbial populations, evolution under these conditions produces a maximization of growth rate~\cite{mazzolini2023universality}. 
However, under non-steady, ecologically realistic conditions, growth rate is not the only contributor to long-term fitness. There is a long tradition of studying the effects of randomly varying conditions on evolutionary dynamics~\cite{levins1968evolution,lewontin1969population,gillespie1973natural,takahata1975effect,levin1984dispersal,frank1990evolution}. Such varying conditions also emerge in explicit ecological contexts, such as predator-prey dynamics~\cite{dobramysl2013environmental} or immune-pathogen coevolution~\cite{weitz2005coevolutionary,marchi2021antigenic}. Phenotypic heterogeneity~\cite{schaffer1974optimal,patra2014phenotypically} and bet-hedging~\cite{kussell2005phenotypic,kussell2005bacterial,acar2008stochastic,beaumont2009experimental} are also extensively studied as evolutionary outcomes of varying environments. These works typically assume unstructured, completely unpredictable environments, where no prediction about future conditions is possible.

On the other hand, organisms cope with varying environments by anticipating change. In the presence of cues, microbial populations can sense the incoming change of conditions and prepare for it. For instance, microbes control their life-cycle to be prepared for changes by sensing changes in nutrients~\cite{liu2015reliable}, cyanobacteria forecast seasonal shifts by measuring light differences~\cite{jabbur2024bacteria}, and bacteria sense nutrient shortage to anticipate stressful conditions~\cite{mukherjee1998shortage}. Cell communication and quorum sensing are fundamental mechanisms allowing cells to anticipate and prepare for imminent nutrient depletion and overcrowding~\cite{goo2012bacterial}. Importantly, microbes anticipate change even in absence of cues, by associating an \textit{a priori} independent behavior. \textit{E. coli} expresses anaerobic metabolic pathways when exposed to a temperature increase~\cite{tagkopoulos2008predictive} and express enzymes and transporters for maltose when growing on lactose~\cite{mitchell2009adaptive}. Analogously, soil bacteria use light cues to anticipate water loss~\cite{hatfield2023light} and growth-rate to predict the need of motility for chemotaxis~\cite{ni2020growth}. By adapting to the temporal order of appearance of environments in their typical habitat, microbes effectively anticipate and prepare for the imminent future. 



Microbial cells encode anticipation through the pre-expression of proteins needed for potential future conditions~\cite{tagkopoulos2008predictive,mitchell2009adaptive,balakrishnan2023conditionally}.
These pre-existing level of proteins expressed for the new condition critically determines the ability of cells to physiologically adapt to the new condition~\cite{basan2020universal,mukherjee2024plasticity,zhu2024shaping}, e.g., affecting the time needed to grow in a new nutrient (lag-time) or the survival to stressful conditions. Preparing for the new condition comes at the cost of growth in the present condition: an increasing proteome fraction associated with future conditions requires a lower fraction to support growth in the present~\cite{balakrishnan2023conditionally,zhu2024shaping}.  Shorter-lag times between conditions are associated to slower growth prior to the switch~\cite{erickson2017global,basan2020universal,mukherjee2024plasticity} and faster growth correlates with lower survival under stress\cite{zhu2023fitness,mukherjee2024plasticity}. Proteome allocation is likely a plastic trait under evolutionary control~\cite{mukherjee2024plasticity,balakrishnan2023conditionally}. Therefore, we can reasonably expect that allocation strikes a balance between current performance and physiological adaptation to future conditions. However, how proteome allocation contributes to the long-term fitness of the population and whether and how it gets shaped by the the temporal structure of the environment is still unclear. In other words, what does the proteome allocation of an organism tell us about the statistical structure of the uncertain environment it evolved in?

Here, by studying the evolutionary dynamics of anticipation in microbes in fluctuating environments, we directly assess the connection between proteome allocation and environmental structure.  
In our framework, proteome allocation encodes the anticipation strategy. 
We develop a mechanistic model that links proteome allocation decisions to both growth rate and lag time, and analyze how natural selection balances these competing demands. We show that, as an output of evolution, proteome allocation encodes information about the long-term statistics of the environment, representing a form of evolutionary learning that allows populations to adapt to partially predictable conditions.

\section{Results}

\subsection{Physiology motivated tradeoff between growth-rate and lag emerging from pre-allocation}

\begin{figure}[tbp]
	\centering
	\includegraphics[width=0.9\textwidth]{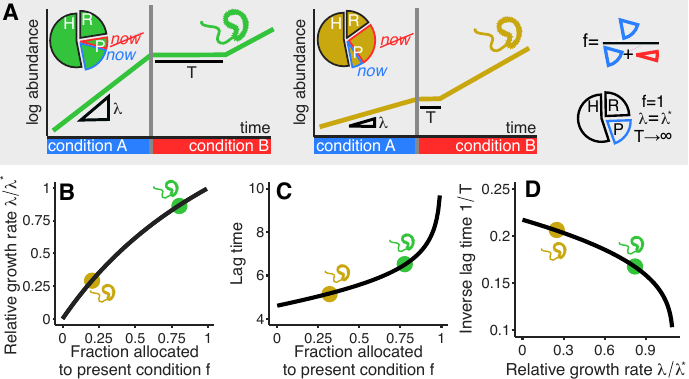} 
	\caption{Proteome partitioning and preallocation. A) Cells allocate their proteome to different functions, which we cluster in three main groups (H: housekeeping, R: ribosomes, P: metabolism). The P-sector is divided into two subparts. A fraction $f$ is allocated to the current condition (condition A, blue) and a fraction $1-f$ to the future condition (condition B, red).
    Cells allocating more to the present condition growth faster in the present condition (panel B), but have a longer lag-time when condition changes (panel C). 
    Taken together, these dependencies on allocation determine a trade-off between growth rate and lag-time (panel D).
    Fig~\ref{sfig:fig1boththeta} generalize the setting to a general class of tradeoff functions. Parameters are reported in the \SI~section~\ref{sisec.sims}.
    }
\label{fig:main1}
\end{figure}

We first focus on cells subject to constant environmental conditions, growing in balanced exponential growth~\cite{jun2018fundamental}. Following a standard approach in microbial physiology~\cite{scott2010interdependence,scott2023shaping,droghetti2025Handson}, we partition their proteome into three main sectors: housekeeping ($H$), ribosome-associated proteins ($R$), and metabolic-associated proteins ($P$), as depicted in Fig~\ref{fig:main1}A. These sectors reflect groups of proteins whose mass fraction have the same linear relationship with growth rate, depending on how conditions are varied~\cite{hui2015quantitative}. 
These three sectors have mass fractions equal to $\phi_H$, $\phi_R$, and $\phi_P$, respectively. 

In the coarse-grained picture of cell processes provided by resource allocation models~\cite{scott2010interdependence}, the sector $P$ is responsible for the mass flux from nutrients to amino acids, while the $R$-sector corresponds to ribosome-associated proteins, assembling amino acids to build proteins. 
Under balanced exponential growth, the population growth rate $\lambda$ equals both~\cite{scott2010interdependence,scott2023shaping}
\begin{equation}
\lambda = \kappa_t \left( \phi_R - \phi_{R,0} \right) = \tilde{\kappa}_n \phi_P   \ ,  
\label{eq.growthlaws}
\end{equation}
where $\kappa_t$ is proportional to the ribosome elongation rate, $\phi_{R,0}$ correspond to inactive ribosomes, and $\tilde{\kappa}_n$ is a nutrient-dependent quantity, often referred to as nutrient quality. The two terms $\tilde{\kappa}_n \phi_P$ and  $\kappa_t \left( \phi_R - \phi_{R,0} \right)$ are the two mass-fluxes (from nutrients to amino acids and from amino acids to proteins, respectively).

Recent work~\cite{balakrishnan2023conditionally,mukherjee2024plasticity} has shown that not all the proteins of the $P$-sector contribute to the flux of mass from nutrients to amino acids. Let us call $f$ the mass fraction of $P$ proteins contributing to that flux. The second term of eq.~\ref{eq.growthlaws} can then be rewritten as $\lambda = \kappa_n f \phi_P$,
where the intrinsic nutrient quality $\kappa_n = f \tilde{\kappa}_n$ captures the biochemical properties of the nutrient the population is growing on, and $f\phi_P$ is the fraction of protein mass devoted to the metabolism of that nutrient. 

The population growth rate $\lambda$ depends on four main quantities: the maximal fraction of active ribosomal protein mass $1-\phi_H -\phi_{R,0}$, the translation capacity $\kappa_t$, the intrinsic nutrient quality $\kappa_n$, and $f$ the fraction of $P$-protein mass devoted to growth under the current conditions. 

The translation speed and the housekeeping proteins set the maximal growth rate achievable $\lambda_{\max}$ under the most optimal condition. This value is achieved when the intrinsic quality of the nutrients is high $\kappa_n \gg \kappa_t$ and growth is limited by the elongation speed of ribosomes alone~\cite{scott2010interdependence}. For finite $\kappa_n$, we define the maximum growth rate $\lambda^*$ corresponding to the case where all the proteins of the $P$-sector are devoted to growth in that condition, i.e., when $f=1$. The value of $\lambda^*$ represent the maximal growth rate under a specific condition. When a fraction of the $P$-sector is devoted to grow in conditions other than the one cells are exposed to ($f<1$), the growth rate will be lower than the maximal $\lambda<\lambda^*$.
Figure~\ref{fig:main1}B shows the dependency of the growth rate on $f$, which reads (see \SI Section ~\ref{si.sec.physiology})
\begin{equation}
\lambda = \frac{f \lambda^*}{1-(1-f) \lambda^* / \lambda_{\max}}   \ .  
\label{eq.grareallocation}
\end{equation}

The growth rate difference between two different nutrients could be due to any combination of intrinsic biochemical differences between the two sugars (different $\lambda^*$) and different allocation strategies (different $f$).
Recent experiments~\cite{mukherjee2024plasticity} have considered the difference between growth rate in mannose and glucose in \textit{E. coli} and shown that the difference is mainly due to the value of $f$, while the values of $\lambda^*$ are remarkably similar. 
Larger values of $f$ correspond to faster growth, but this advantage comes at the cost of slower physiological adaptation to new changing conditions (under the form of longer lag or increased mortality)~\cite{mukherjee2024plasticity}. 

In this work, we focus on nutrient flux as the main bottleneck for growth under changing conditions. We consider the scenario when a population growing under a given condition is suddenly exposed to a new condition. We assume that this new condition is associated with metabolic pathways corresponding to a fraction of protein mass $\phi_{\new}$. Immediately following the switch of conditions, the value of $\phi_{\new}(t)$ increases over time to eventually converge to the value observed in balanced exponential growth $\bar{\phi}_{\new}$. We assume that the dynamics of $\phi_{\new}(t)$ follows~\cite{erickson2017global}
\begin{equation}
\frac{d \phi_{\new}}{dt} = \frac{ \bar{\phi}_{\new} }{\tau} h\left( \frac{\phi_{\new}(t)}{\bar{\phi}_{\new}} \right) \left( 1 - \frac{\phi_{\new}(t)}{\bar{\phi}_{\new}} \right) \ .  
\label{eq.sectordyn}
\end{equation}
The term $1 - \phi_{\new}(t) / \bar{\phi}_{new}$ ensures the convergence of $\phi_{\new}(t)$ to $\bar{\phi}_{\new}$. The other terms capture the speed of growth of $\phi_{\new}(t)$ during the phase of physiological adaptation.
We assume that when a nutrient changes, the availability of amino acids to build the proteins needed to metabolize the new nutrient is limited by those same proteins that contribute to this flux.
In particular, the monotonically increasing function $h(\cdot)$ captures this non-trivial dependency, and the parameter $\tau$ is the associated timescale. 
In the following, we consider the dependency $h(z) = z$. The rate of production of new enzymes if proportional to the ribosomal flux. Assuming that ribosomal and nutrient flux are matched~\cite{erickson2017global} one obtains that, for small $\phi_{\new}$, the rate of increases of the new protein sector is proportional to the nutrient flux ($d \phi_{\new} / dt \propto \phi_{\new}$).
In the \SI, we consider explicitely the more general case $h(z)=z^\theta$.
In particular, the case $\theta = 2$ has biological relevance, corresponding to the switch from a glycolitic to gluconeogenic carbon source~\cite{basan2020universal}.

We assume that the growth rate is limited by the metabolism of the new nutrient, with a corresponding time-dependent growth rate $\mu(t) = \kappa_n \phi_{\new}(t)$. 
For large times, the growth rate converges to the value $\lim_{t \to \infty} \mu(t) = \lambda = \kappa_n \bar{\phi}_{\new}$ corresponding to equation~\ref{eq.growthlaws}. A population growing with a time-dependent growth rate will have abundance $x(t)$ at time $t$ given by
$\log x(t) / x(0) = \int_0^t ds \  \mu(s)$. For large times, one can effectively capture this dependency by introducing a lag time $T$ defined as
$\log x(t) / x(0) = \lambda ( t- T)$. Using this definition and equation~\ref{eq.sectordyn}, with the choice $h(z) = z$, one obtains (see \SI)
\begin{equation}
   T = - \tau \log \left( \frac{\phi_{\new}(0)}{\bar{\phi}_{\new}} \right) \ .
   \label{eq:lagtime1}
\end{equation}

According to eq.~\ref{eq:lagtime1} the lag time is therefore a decreasing function of the initial allocation $\phi_{\new}(0)$: this initial allocation is what limits the growth in the initial phase following the condition change setting the time needed to adjust the proteome. Importantly, this initial allocation is determined by the expression levels in the previous (old) condition. The value $\phi_{\new}(0)$ is in fact equal to $q(1-f_{\old})\phi_P$, where $(1-f_{\old})\phi_P$ is the mass fraction pre-allocated (i.e., allocated before the switch) to proteins needed for all possible conditions other than the old one, while $q$ is the fraction, among those, allocated for growth in the new condition. On the other hand, one has $\bar{\phi}_{\new} = f_{\new} \phi_P$ Assuming that the new and old conditions have the same nutrient quality $\kappa_n$, by substituting this expression in eq.~\ref{eq:lagtime1}, one obtains an explicit dependency of the lag time on the allocations $f_{\old}$ and $f_{\new}$. The lag-time is an increasing function of both $f_{\old}$ and $f_{\new}$. Fig.~~\ref{fig:main1} displays the dependency of the lag-time on the allocation $f_{\old}$. If the values of $f$ are expressed in terms of the growth rate using eq.~\ref{eq.grareallocation}, one obtains
\begin{equation}
   T = 
 - \tau \log \left( q \frac{\lambda^*-\lambda_{\old}}{\lambda_{\new}} \left(1 -\frac{\lambda^*}{\lambda_{\max}}\right)\right)
   \ .
   \label{eq:lagtime2}
\end{equation}
which makes explicit the trade-off between growth rate and lag-time~\cite{erickson2017global,basan2020universal}, displayed in
Fig.~\ref{fig:main1}D.


\subsection{Evolutionary stable state in boom-and-bust environment with deterministic switching}

We showed how increasing the mass fraction of proteome allocated to the present condition $f$ increases the growth rate at the expenses of a longer lag-time. We expect the allocation $f$ to be under selection pressure and change evolutionary time. We expect the environmental context (e.g., the rate at which conditions change) to determine an evolutionary balance between faster growth and shorter lag~\cite{manhart2018trade,mukherjee2023coexisting}.

\begin{figure}[tbp]
	\centering
	\includegraphics[width=0.55\textwidth]{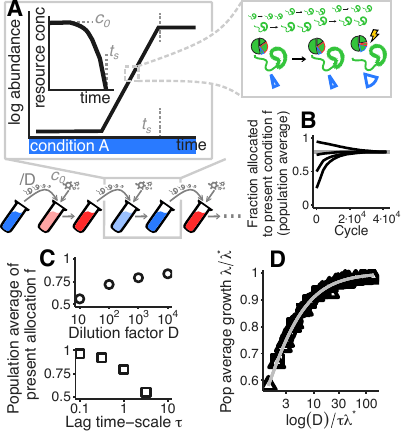} 
	\caption{
    Evolution of allocation parameter in serial dilution. We consider a population evolving in a boom-and-burst environment (panel A). Cells are added to fresh medium in condition A and start growing exponentially after a lag-time and until the nutrients, originally present in concentration $c_0$, are depleted (at time $t_d$). Cells are then diluted, with a constant factor $D$, to fresh medium in condition B. Conditions alternate deterministically over cycles. During duplications cells are subject to mutation that alter the fraction of proteome $f$ allocated to the present condition. The fate of these mutations is determined over (a large number of) cycles. Panel B shows that the population average of $f$ converge to a final value independently of the allocation fraction $f$ of the initial clonal population. 
    Panel C shows the dependency of this final value of $f$ on the parameters: it increases with the dilution factor $D$, while it decreases with the lag time-scale $\tau$. Evolution drives allocation to a value that balances the effect of population growth and lag-time, resulting in a growth rate $\lambda$ lower than the limiting one $\lambda^*$ (corresponding to the case $f=1$).
    Panel D shows that the evolved value of the relative growth rate $\lambda / \lambda^*$ depends on the adimensional quantity $\log D/ (\lambda^* \tau)$. The gray line corresponds to the prediction of eq.~\ref{eq.lambdaess}.
    Parameters are reported in the \SI~section~\ref{sisec.sims}.
    }
\label{fig:main2}
\end{figure}

We start by considering a boom-and-bust environment with two alternating conditions (e.g., two different nutrients, see Fig.~\ref{fig:main2}A). Nutrients are provided at constant concentration $c_0$ and the population grows until resources are depleted. Populations are diluted by a constant factor $D$, and exposed to the other condition, with nutrients also provided at concentration $c_0$. The process is repeated over dilution cycles by switching between the two conditions.

We assume that the allocation value $f$ is an evolvable trait and ask to what value(s) an evolving population converges to. Each individual within the populaton is characterized by a value of $f$, which we assume is genetically encoded and that determines the growth rate and lag-time. When an individual reproduces, mutation can occur and the offspring can have a slightly different value of $f$. 
We define a strain, indexed by $i$, as the group of individuals with a given value of allocation $f_i$.
The population average allocation parameter $f$ at the $d$-th cycle is defined as
$\langle f \rangle(d) := \sum_i x_i(d) f_i$
where $x_i(d)$ is the relative abundance of strain $i$ with value $f_i$ at the end of the $d$-th cycle.
Figure~\ref{fig:main2}B shows that the population-averaged value $\langle f\rangle $ of a population converges to the same value regardless of the initial condition. Moreover, Figures~\ref{sfig:varianceevolving1} and~\ref{sfig:varianceevolving2} show that the population variance of $f$ --- defined as $\sigma_f^2(d) = \sum_i x_i(d) f_i^2 - \langle f \rangle^2$ --- converges to low values. Together, these two observations indicate that evolutionary dynamics drive the population to an evolutionary stable value $f^{\ess}$ that does not depend on the initial condition.

The value $f^{\ess}$ represents an optimal balance between growth and lag. This optimal balance is context-dependent and experimentally tunable. Fig.~\ref{fig:main2}B shows that larger dilution factors $D$ correspond to larger $f^{\ess}$. Analogously, larger values of $\tau$ correspond to lower $f^{\ess}$.

Since the outcome of evolution is context dependent (e.g., depending on the dilution factor $D$), one could ask what ultimately determines the evolutionary stable value $f^{\ess}$. Let us imagine that one could measure all the growth properties (growth rate and lag time, determined by $f$) of a strain in isolation. Which properties determine whether that strain will outcompete another one? To answer these questions, we employ invasion analysis (see Supplementary Information~\ref{secsi.invationsym}) to quantify the invasion growth rate of an invading strain $j$ with allocation value $f_j$ when the resident strain $i$ has allocation value $f_i$. The answer is remarkably simple: when competing, the strain that exhibits the shortest depletion time in isolation always outcompetes the other. In isolation, the resource depletion time $t_s(f)$ of a strain with allocation value $f$ is given by
\begin{equation}
 t_s(f)  = T(f) + \frac{ \log D }{\lambda(f)} \ .
\end{equation}
The strain $j$ will invade the resident strain $i$ if $t_s(f_j) < t_s(f_i)$.
This result implies that there is one evolutionary stable state corresponding to the value of $f^{\ess}$ that minimizes the resource depletion time
\begin{equation}
f^{\ess} = \min_{f} t_s(f) \ .
\end{equation}
In general, $f^{\ess} < 1$, which corresponds to a value of the growth rate lower than the maximum achievable in that condition $\lambda(f^{\ess}) < \lambda^*$. The resulting expression for the evolutionary stable value of the growth rate reads
\begin{equation}
\lambda(f^{\ess}) = \lambda^* \left( 1 +  \frac{\tau \lambda^*}{\log D}   \right)^{-1}  \ .
\label{eq.lambdaess}
\end{equation}
Fig.~\ref{fig:main2}D shows that the end state of the simulated evolutionary trajectories converge to the predicted value.

In the case of more general trade-off functions (corresponding to $h(z)=z^{\theta}$ in eq.~\ref{eq.sectordyn}, we obtain a similar result (see \SI~\ref{si.sec.physiology.lag}). The value of the evolutionary stable growth rate has the form $\lambda(f^{\ess}) = \lambda^* \alpha/(1+\alpha)$ where $\alpha$ is a dimensionless variable capturing the value of the parameters and the shape of the tradeoff. 
In particular, $\alpha$ depends on the ratio between two time scales: $\tau$ (the time-scale associated with physiological adaptation and lag-time, see eq.~\ref{eq.sectordyn}) and $\log D / \lambda^*$ (the time spent in the growth phase by a population growing of a factor $D$ with maximum growth rate $\lambda^*$). The ratio between these two time scales strikes the balance between growth rate maximization and lag-time minimization.

\subsection{Evolutionary stable state in stochastic boom-and-bust environment}

The previous section considered the case where resource supply $c_0$ and dilution factor $D$ have always the same value. Here we consider a stochastic environment where the dilution factor $D(d)$ and resource supply $c_0(d)$ vary across cycles.
Moreover, we consider $M > 2$ conditions: after growth under a given condition, each of the $M-1$ other condition can potentially occur.
These choices make the environment stochastic, as we
consider that $D(d)$ and $c_0(d)$ are random variables and that each of the $M-1$ conditions can follow any other with equal probability $1/(M-1)$.

\begin{figure}[tbp]
	\centering
	\includegraphics[width=\textwidth]{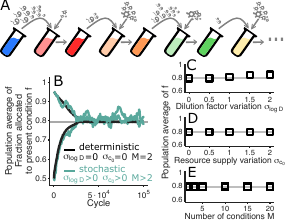} 
	\caption{Evolution of allocation in a variable environment. Panel A). We consider a boom-and-bust setting where the amount of supplied nutrient $c_0$ and the dilution factor $D$ vary across cycles. We also consider $M>2$ conditions, making the sequence of conditions stochastic.
    Panel B). The evolutionary trajectories display larger noise, but still converge to a unique value of $f$, compatible with the deterministic case observed in Fig.~\ref{fig:main2}. 
    The simulated evolved value of $f$ (points) does not depend on the variability of the dilution factor (quantified by $\sigma_{\log D}$, the variance of $\log D$, panel C), the variability of the resource supply (quantified by the variance $\sigma_{c_0}$, panel D), and the number of condition $M$ (panel E).
    Parameters are reported in the \SI~section~\ref{sisec.sims}.
    }
\label{fig:main3}
\end{figure}

Fig~\ref{fig:main3}A shows two examples of evolutionary trajectories of the population average of the allocation value $\langle f \rangle$, under a stochastic environment. Although stochasticity determines fluctuations of the allocation value over cycles, these values oscillate around the value $f^{\ess}$ observed for deterministic environments.

In the deterministic case, the invasibility criterion allowed us to derive the existence of a unique ESS corresponding to the value of $f$ minimizing the time of resource depletion. In the stochastic case, the depletion time is a random variable, as it depends on resource supply $c_0$, on dilution factor $D$, and on which of the $M$ conditions the population is exposed to.

In this section, we focus on the simpler case where the $M$ conditions are statistically equivalent ($c_0$ and $D$ are independent and are both identically distributed across conditions). 
Surprisingly, we show that the invasibility condition is particularly simple: a strain will invade a resident if it has a lower \textit{expected} depletion time (see \SI~\ref{secsi.invasionfull}).
Fig.~\ref{fig:main3}C, D and E, shows that the average depletion time is independent of the fluctuation of $c_0$ and $D$, and on the number of conditions $M$. Equation~\ref{eq.lambdaess} generalizes to this case as
\begin{equation}
\lambda(f^{\ess}) = \lambda^* \left( 1 +\frac{\tau \lambda^*}{\mathbb{E}( \log D ) }  \right)^{-1}  \ .
\label{eq.lambdaessfluct}
\end{equation}
In particular, the ratio between the lag time-scale $\tau$ and the expected growth phase duration at maximum growth $\langle \log D \rangle / \lambda^*$ still controls the value of the evolutionary stable growth rate. These results apply also for other trade-off shapes, except for the independence of eq.~\ref{eq.lambdaessfluct} on the number of conditions, which can affect the value of the evolutionary stable allocation for more general tradeoff shapes (see \SI, Figure~\ref{sfig:fig2boththeta}).


\subsection{Evolutionary learning: optimal pre-allocation matches statistic of the environment}

In the previous section, we considered an environment where each condition was equally likely to follow another one. In this section, we focus on a more complex environmental structure, where sequences of conditions are not equally probable. We encode this dependency in a transition matrix $\mathbf{w}$, whose elements $w_{a|b}$ quantify the probability that condition $b$ is followed by condition $a$. 

\begin{figure}[tbp]
	\centering
	\includegraphics[width=\textwidth]{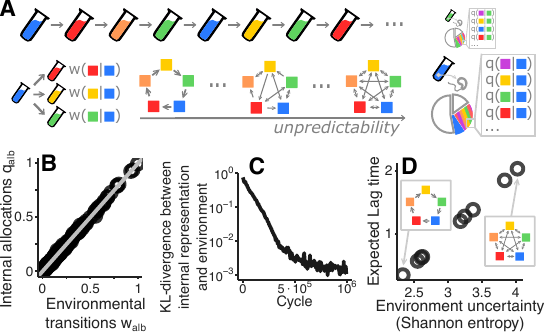} 
	\caption{Evolution in a partially predictable environment. We consider a partially predictable environment, where a sequence of condition is drawn from a Markov process specified by a transition matrix $\mathbf{w}$ (panel A). In parallel, the fraction of proteome $1-f$ allocated to condition different from the present one, is further portioned across the potential future conditions. The matrix $\mathbf{q}$ encodes this allocation. 
    Panel B shows the outcome of the evolutionary simulations where $\mathbf{q}$ and $f$ are allowed to evolve. The evolved values of $\mathbf{q}$ (colored points) match the transition probability as predicted by adaptive dynamics. Fig~\ref{sfig:fig4theta} generalize these results to other tradeoff shapes and levels of uncertainty of the environment.
    Panel C shows that the evolutionary dynamics minimizes the Kullback-Leibler divergence between $\mathbf{w}$ and $\mathbf{q}$.
    Panel D reports the value of the expected lag-time for the evolved allocation $\mathbf{q} \sim \mathbf{w}$ for different choices of the transition matrix $\mathbf{w}$. The expected lag-time displays a linear dependency with the transition dynamics uncertainty (quantified by the Shannon entropy).
    Parameters are reported in the \SI~section~\ref{sisec.sims}.
    }
\label{fig:main4}
\end{figure}

Since we break the symmetry between condition, it is worth introducing the variable $\phi_{a|b}$ that quantifies the mass fraction of the proteome allocated to proteins needed for condition $a$ while growing in condition $b$. By definition, the total amount allocated is equal to the size of the $P$-sector during growth in condition $b$: $\phi_{P,b} = \sum_{a} \phi_{a|b}$. 
It is convenient to write (see Fig.~\ref{fig:main4}A)
\begin{equation}
    \phi_{a|b} = \phi_{P,b} \left(
f_b \delta_{a,b} + (1-f_b) (1-\delta_{a,b}) q_{a|b} 
\right) \ ,
\label{eq.phidecomp}
\end{equation}
where $\delta_{a,b}$ is the Kronecker delta.
The entry $f_b$ of the vector \underline{f} quantifies the mass fraction of the $P$-sector allocated to growth in condition $b$ while growing in condition $b$. The fraction $1-f_b$ is the one allocated to conditions other than $b$ while growing in $b$. The entry $q_{a|b}$ of the matrix $\mathbf{q}$ specifies how much of this fraction is (pre-)allocated to condition $a$, with the constraint $\sum_a q_{a|b} = 1$ and the prescription $q_{a|a}=0$.
Equation~\ref{eq.phidecomp} naturally splits the proteome allocation into two components: the values of $\underline{f}$, which determine how much to allocate to the present condition vs. unknown future ones, and the values of $\mathbf{q}$, which determine proportionally how much to allocate to each of the unknown future conditions.

In the previous sections, we considered an unstructured environment, with uniform $w_{a|b} = 1/(M-1)$, and symmetric conditions, which naturally corresponded to an unstructured partitioning $q_{a|b} = 1/(M-1)$ and symmetric allocations $f_b  = f^{\ess}$ independent of $b$.
When the transitions are not uniform ($w_{a|b} \neq 1/(M-1)$), we expect that the evolutionary stable solution  do not correspond to the uniform solution $q_{a|b} = 1/(M-1)$. In fact, when the transitions between conditions are not equally probable, it might be more advantageous to have non-uniform pre-allocations encoded in $\mathbf{q}$. For instance, consider the extreme case when condition $c$ never follows condition $a$. In this case pre-allocating to condition $c$ while growing in $a$ ($q_{c|a}>0$) will not produce any advantage to the growing cells, while sequestering resources that could be used otherwise.

As expected, the value of $\mathbf{q}$ contributes significantly to the adaptation in a random environment. 
The evolutionary dynamics drives the population to an unique evolutionary stable value of $\mathbf{q}^{\ess}$ that, by minimizing the expected depletion time, matches the statistical structure of the environment defined by the transition matrix $\mathbf{w}$ with the simple condition
\begin{equation}
q_{a|b}^{\ess} =  w_{a|b}
\ .
\label{eq.qess}
\end{equation}
Figure~\ref{fig:main4}B shows this relationship for the endpoint of the population average $\langle \mathbf{q} \rangle$ of the simulated evolutionary trajectories.
The simple relationship of eq.~\ref{eq.qess} holds because, for the tradeoff shape of eq.~\ref{eq:lagtime2} the contribution of the allocation $\underline{f}$ and the pre-allocation $\mathbf{q}$ become independent (see \SI section~\ref{secsi.adaptiveboom}). While a non-uniform transition matrix $\mathbf{w}$ corresponds to conditions occurring with different probability, the allocations $f_a$ evolve to the same value $f^{\ess}$ (see Figure~\ref{sfig:mean_f_time} and~\ref{sfig:sd_f_time}). We explicitly discuss the result for other shapes in \SI~section~\ref{secsi.adaptiveboom}.

For large times, the population average $\langle \mathbf{q} \rangle$ converges to $\mathbf{q}^{\ess} = \mathbf{w}$. In this state, the fraction of the proteome allocated to a given potential future condition exactly matches the probability that it will occur.
In the \SI~\ref{secsi.adaptiveboom}, we use adaptive dynamics~\cite{metz1995adaptive,diekmann2002beginners} (see \SI~\ref{secsi.adaptiveprimer} for a short introduction on adaptive dynamics) to derive an equation for the evolutionary trajectories leading to this final state. We show that the population average $\langle \mathbf{q} \rangle$ of the pre-allocation values follows the gradient of the Kullback-Leibler divergence between itself and the transition probabilities:
\begin{equation}
\frac{ \partial \langle q_{a|b} \rangle }{\partial t} \propto
- \frac{\partial }{ \partial \langle q_{a|b} \rangle } D_{KL}(\mathbf{q}|\mathbf{w}) + \text{n.c.}
\ ,
\label{eq.KLdiv}
\end{equation}
where we do not write explicitly the normalization conditions (n.c.) ensuring $\sum_a q_{a|c}=1$.
This correspondence emerges from the logarithmic dependency of eq.~\ref{eq:lagtime2}, but it nevertheless determine a mathematical equivalence between evolutionary dynamics and learning.

The evolutionary stable state $\mathbf{q}^{\ess}$ minimizes the expected depletion time. Since $\mathbf{q}$ does not influence the growth rate, its value corresponds to the one minimizing the expected lag-time. This expectation is calculated over the uncertainty of the sequence of conditions, defined by the matrix $\mathbf{w}$. 
\begin{equation}
 \mathbb{E}(T(\mathbf{q}^{\ess})) = c - \sum_a p_a \sum_b w_{b|a} \log w_{b|a}
\ ,
\label{eq.Tlafmin}
\end{equation}
where $c$ is a function of the dilution factor $D$ and $\underline{f}^\ess$, independent of $\mathbf{w}$.
The expected lag-time depends on the transition matrix through a term equal to the Shannon entropy of the transition matrix $\mathbf{w}$ (see \SI~\ref{sisec.learning} and Fig.~\ref{fig:main4}D).  Eq.~\ref{eq.Tlafmin} directly quantifies the cost of the unpredictability of the environment. The more the conditions are unpredictable (larger entropy values), the longer is the expected lag time: the amount of time lost to growth due to the unpredictability is directly proportional to the bits needed to describe the statistics of the dynamic environment.

\subsection{Evolution under non-symmetric conditions}

The previous sections focus on the symmetric case, where the evolutionary dynamics drive the system to an evolutionary stable state of allocation equal across all conditions: $f_a^\ess = f^\ess$ for all conditions $a$. In this case the growth rate is equal across conditions: $\lambda_a = \lambda(f^{\ess})$. 

Using adaptive dynamics, in section~\ref{secsi.adaptiveboom} we derive the general equations for the evolution of the population average growth rates $\langle \underline{\lambda} \rangle$ corresponding to the allocations $\underline{\lambda}$, which read
\begin{equation}
\frac{d \langle \lambda_a \rangle}{d t} = -
\sum_c p_c \langle \lambda_c \rangle \frac{\partial }{\langle \lambda_a \rangle}
\mathbb{E}_c\left( t_{s}(\langle \underline{\lambda} \rangle, \langle \mathbf{q} \rangle)\right)
\ ,
\label{eq.adptf}
\end{equation}
where $\mathbb{E}_a\left( t_{s}(\langle \underline{\lambda} \rangle, \langle \mathbf{q} \rangle)\right)$ is the expected value of the depletion time under condition $a$ and $p_a$ is the probability that condition $a$ occurs. An analogous equation defines the dynamics of $\langle \mathbf{q} \rangle$. These equations define a non-linear dynamical system describing the evolutionary trajectories. In general, if conditions are symmetric --- more specifically, if $\mathbb{E}_a\left( t_{s}(\langle \underline{\lambda} \rangle, \langle \mathbf{q} \rangle)\right)$ is condition independent --- there exists a unique fixed point (i.e., an evolutionary stable state) corresponding to the minimum of the expected depletion time.

This symmetry condition can be broken in multiple ways. Under a general transition matrix $\mathbf{w}$, condition $a$ will occur with probability $p_a$, determined by the stationarity condition $p_a = \sum_b w_{a|b} p_b$. Under the tradeoff of eq.~\ref{eq:lagtime2} the fixed points of eq.~\ref{eq.adptf} are still symmetric, independently of $p_a$. For other tradeoff shapes however, this simplification breaks and both $\underline{f}^{\ess}$ (and the corresponding growth rates $\underline{\lambda}^{\ess}$) become condition dependent. Section~\ref{secsi.adaptiveboom} derives the set of equation for the evolutionary stable state in the general case. Figures~\ref{sfig:mean_f_time} and~\ref{sfig:fig4theta} that the (pre)-allocation minimizing the expected depletion time provide a good approximation to the end-point of the evolutionary dynamics.

Another way to break the symmetry condition is to assume that the distributions of the dilution factor $D$ and of the resource supply $c_0$ are condition dependent. In this case, for the tradeoff eq.~\ref{eq:lagtime2}, we obtain in section~\ref{secsi.adaptiveboom} that the evolutionary stable growth rates are the solution of the equation
\begin{equation}
\lambda_a^\ess = \lambda^* \left(
1 + \frac{\tau \lambda^*}{\log \bar{D}_a } \left(
1 + \sum_a w_{b|a} \frac{\lambda_b -\lambda_a}{\lambda^*} 
\right)
\right)^{-1}
\ ,
\label{eq.adpsollambda}
\end{equation}
where $\log \bar{D}_a$ is the expected log growth factor (logarithm of the final over initial biomass) of a growth cycle under condition $a$. Under symmetric conditions $\log \bar{D}_a = \mathbb{E}(\log D)$ and eq.~\ref{eq.adpsollambda} reduces to eq.~\ref{eq.lambdaessfluct}. If this symmetry is broken, the values of $\underline{\lambda}^\ess$ depend on the differences in growth factor, but also on the transition rates. In particular, the term $\sum_b w_{b|a} (\lambda_b -\lambda_a) / \lambda^* $ quantifies the difference between the expected growth in the future and the growth in the presence. The larger is this difference (the more the future is expected to be favorable) the less cells will allocate for the present condition.

The preallocation values $\mathbf{q}^\ess$ are also affected by the difference between conditions. Eq.~\ref{eq.qess} generalizes to 
\begin{equation}
q_{a|b}^{\ess} = \frac{ \lambda_a^\ess w_{a|b} }{ \sum_c \lambda_c^\ess w_{c|b} }
\ .
\label{eq.qessgen}
\end{equation}
The evolved preallocation values weight the likelihood of the alternative future conditions $w_{a|b}$ with the growth rate in those condition. The larger is the growth rate the larger pre-allocation will be.
Section~\ref{secsi.adaptiveboom} reports the generalization of eq.~\ref{eq.adpsollambda} and~\ref{eq.qessgen} for arbitrary trade-off shapes.

\section{Discussion}

Our results show how microbial populations facing environmental uncertainty must navigate a fundamental trade-off between specialization for current conditions and preparation for future ones. The evolutionary stable allocation value $f^{\ess}$ represents an optimal balance that minimizes resource depletion time by weighing the benefits of faster growth against the costs of longer adaptation times. This balance emerges from the specific temporal structure of the environment, and in particular from the ratio between physiological adaptation timescales ($\tau$) and growth phase duration ($\log D / \lambda^*$). When environmental switches are frequent relative to adaptation time, selection favors reduced allocation to current conditions ($f^{\ess} < 1$), sacrificing some growth rate to gain faster transitions. Uncertainty itself becomes a selective pressure that shapes cellular resource allocation strategies.

Perhaps most remarkably, we show that populations can evolutionarily ``learn'' the statistical structure of their environment through proteome pre-allocation patterns. The evolved pre-allocation matrix $\mathbf{q}^{ess}$ converges to values that reflect the transition probabilities between environmental states, effectively encoding environmental predictability in the cellular proteome. The evolutionary dynamics leading to this stable allocation is mathematically equivalent to statistical learning algorithms. The environmental unpredictability directly translates into increased lag times and reduced fitness. This demonstrates that even in the absence of explicit memory or prediction mechanisms --- like the ones of neural systems --- evolutionary processes can extract information and exploit environmental patterns. The proteome serves as a distributed information storage system about environmental statistics.
Our work can be viewed through the broader lens of information theory and its role in biological systems~\cite{donaldson2010fitness,pinero2024information}. Our results are an explicit realization of information-limited growth, which should obey information-theoretic bounds on productivity or replicators~\cite{pinero2024information}.

Under symmetric conditions, the outcomes of evolutionary dynamics also display a deep analogy with optimal portfolio theory through the Kelly criterion~\cite{kelly1956new}. In this case, in fact, the evolved pre-allocation exactly matches environmental transition probabilities ($\mathbf{q}^{ess} = \mathbf{w}$), analogous to the Kelly strategy in finance where optimal investment fractions equal outcome probabilities. This equivalence suggests that certain physiological constraints lead to allocation strategies that are mathematically optimal for maximizing long-term growth in stochastic environments. 

While our model primarily focuses on allocation-driven differences in growth rates, real environments exhibit additional complexity through variations in intrinsic nutrient quality ($\lambda^*$). Different sources of essential nutrients, pH, temperature and other environmental factors can substantially alter the maximum achievable growth rate. 
The interplay between allocation evolution and adaptation to different intrinsic nutrient qualities can potentially lead to more complex evolutionary dynamics and the maintenance of genetic diversity within populations. This more complex evolutionary dynamics may prevent convergence to a single evolutionary stable state, instead favoring the coexistence of multiple strains with complementary allocation strategies, each specialized for different aspects of the environmental variability.

An important assumption of our work is not to consider phenotypic heterogeneity. Bet-hedging~\cite{schaffer1974optimal,patra2014phenotypically,kussell2005phenotypic,kussell2005bacterial,acar2008stochastic,beaumont2009experimental} is an important outcome of evolution in uncertain conditions, which allows populations to better adapt to variable environments. Our setting serves as a baseline of the maximum level of adaptation achievable by a phenotypically homogeneous population.

Another important assumption of our framework is to consider evolution at the level of proteome allocation. These allocation values are controlled by a complex --- yet limited --- regulatory network~\cite{tagkopoulos2008predictive}. We are assuming that mutations affecting this network can achieve the $M^2$ arbitrary values of $\underline{f}$ and $\mathbf{q}$. The complexity of regulatory networks required to achieve complex allocation patterns come at a cost, we should be accounted as an important constraint of evolution (e.g., also affecting the emergence of phenotypic heterogeneity~\cite{kussell2005phenotypic}).

While our analysis focuses on the growth rate-lag time trade-off, real cellular populations face additional constraints that complicate allocation strategies. Survival during starvation periods, resistance to environmental stresses, and maintenance of cellular integrity all depend on proteome allocation and may conflict with strategies optimized solely for growth and lag time. For instance, allocating resources to stress response proteins or maintenance functions might further reduce growth rates~\cite{zhu2023fitness} while providing survival benefits during harsh transitions. These multi-dimensional trade-offs suggest that natural populations operate under even more constrained optimization problems, where allocation strategies must balance growth, adaptation speed, stress resistance, and survival probability simultaneously, potentially leading to more conservative allocation strategies than predicted by growth-lag trade-offs alone.

Our work exemplifies how information processing and learning can emerge from simple evolutionary dynamics without requiring neural networks or explicit computational machinery. The proteome allocation system functions as a distributed memory device that stores information about environmental patterns and uses this information to make predictive ``decisions'' about resource allocation. This represents a form of embodied cognition where the physical structure of the organism (protein expression levels) directly encodes learned environmental patterns. The mathematical equivalence between evolutionary
dynamics and gradient descent on information-theoretic objectives suggests that such learning mechanisms may be widespread in biology, from gene regulatory networks that anticipate circadian rhythms to immune systems that maintain memory of past infections. This framework provides a foundation for understanding how organisms can exhibit apparently intelligent behavior through the evolution of their molecular and cellular architecture, demonstrating that learning need not require centralized computational systems but can emerge from the  evolution of biological replicators.


\section{Acknowledgments}

We thank S. Azaele, A. Bupu, L. Fant, A. Goyal, S. Maslov, M.A. Mu\~noz, E. Pigani, M. Sireci, S. Suweis, W. Shoemaker for insightful discussions.
O.M. acknowledges the Trieste Laboratory on Quantitative Sustainability - TLQS for funding.

\bibliographystyle{naturemag}

\clearpage

\clearpage


\begin{appendix}

\clearpage

\clearpage
\begin{center}
    {\LARGE Supplementary Material} \\
\vspace{0.2 in}
\thispagestyle{empty}
\end{center}

\counterwithin{equation}{section}
\counterwithin{figure}{section}
\counterwithin{table}{section}


\renewcommand{\theequation}{S\arabic{equation}} 
\renewcommand{\thefigure}{S\arabic{figure}}   
\renewcommand{\thetable}{S\arabic{table}}     
\renewcommand{\thepage}{S\arabic{page}} 

\setcounter{page}{1}

\section{Physiology and proteome allocation}
\label{si.sec.physiology}

\subsection{Effect of proteome pre-allocation on growth rate}
\label{si.sec.physiology.prealloc}

A cell allocates its internal resources across multiple functions. By coarse-graining over these functions, one can define sectors that group proteins contributing to similar biological processes that change similarly as a function of the conditions.

We consider a ribosomal sector (quantified by a mass fraction $\phi_R$), a metabolism sector (with a corresponding mass fraction $\phi_P$), and a housekeeping sector with mass fraction $\phi_H = 1-\phi_{\max}$.
Following the setting of resource allocation models~\cite{scott2010interdependence,scott2023shaping,droghetti2025Handson}, we write the growth rate as
\begin{equation}
\lambda = \kappa_t \phi_R  = \tilde{\kappa}_n \phi_P \ ,
\label{sieq.fluxmatch}
\end{equation}
where $\kappa_t$ is the translational efficiency, and $\tilde{\kappa}_n$ as nutrient quality.
Contrarely to the typical approach~\cite{droghetti2025Handson}, in order to simplify the notation, we absorbed the inactive ribosome $\phi_{R,0}$ in the housekeeping sector and we interpret $\phi_R$ appearing in eq.~\ref{sieq.fluxmatch} as the active ribosome. 
The standard notation~\cite{scott2010interdependence,droghetti2025Handson} can be obtained by substituting $\phi_{\max} \to \phi_{\max} - \phi_0$ and $\phi_R \to \phi_R - \phi_0$ in all the expressions below.

Using $\phi_R + \phi_P = \phi_{\max}$, we obtain
\begin{equation}
\lambda = \phi_{\max} \left( \frac{1}{\kappa_n} +  \frac{1}{\kappa_t}  \right)^{-1}  \ ,
\end{equation}
Under optimal conditions ($\kappa_n \gg \kappa_t$) cells grow close to a maximum growth rate 
\begin{equation}
    \lambda_{\max} =  \phi_{\max}  \kappa_t.
\end{equation}

\begin{figure}[tbp]\centering
\includegraphics[width=\textwidth]{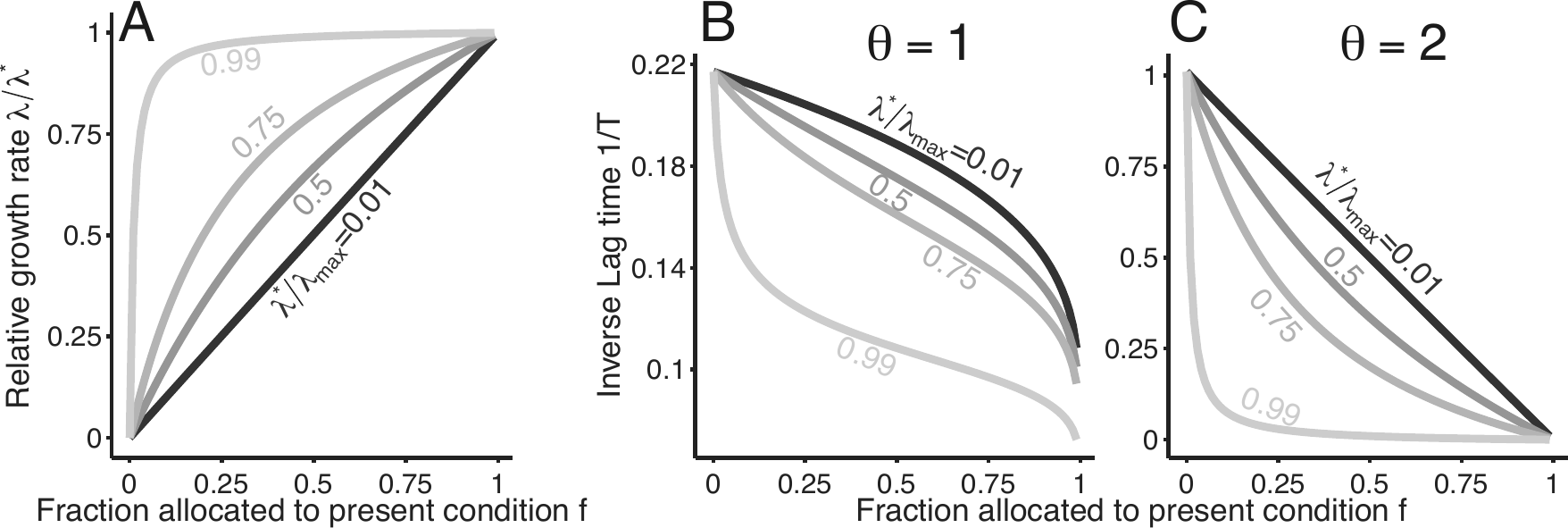} 
\caption{A) Dependency of the growth rate $\lambda(f)$ (relative to the condition maximum $\lambda^*$) as function of the allocation $f$ for different values of $\lambda^*/\lambda_{\max}$ (gray scale), as defined in equation~\ref{si.eq:grate_f}.
B) and C) Inverse lag-time as function of the allocation fraction $f$ for different values of $\lambda^*/\lambda_{\max}$ (gray scale) and two tradeoff shapes ($\theta=1$ in panel B and $\theta = 2$ in panel C).}
\label{sfig:lambdaTvsf}
\end{figure}

Following recent experimental results~\cite{mukherjee2024plasticity}, we assume that the $P$ sector is further divided in two sub-parts $\phi_{P,\now}$ and $\phi_{P,\notnow}$, with $\phi_{P,\now}+\phi_{P,\notnow} = \phi_P$. The subsector, $\phi_{P,\now}$ is the one actually contributing to growth in the current condition cells are experienced. Proteins belonging to $\phi_{P,\notnow}$ (referred to as the adaptability sector~\cite{mukherjee2024plasticity,droghetti2025Handson} are not currently contributing to growth, but they could become useful in a potentially new condition. 
By introducing the parameter $f$ such that $\phi_{P,\now}=f \phi_P$ (and $\phi_{P,\notnow} = (1-f) \phi_P$), we obtain
\begin{equation}
\lambda = \phi_{\max} \left( \frac{1}{f\kappa_n} +  \frac{1}{\kappa_t}  \right)^{-1}  \ .
\label{sieq:allocationfluxes}
\end{equation}
If cells allocate fully for the present condition ($f=1$) they would grow with growth rate $\lambda^*$ given by
\begin{equation}
\lambda^* = \phi_{\max} \left( \frac{1}{\kappa_n} +  \frac{1}{\kappa_t}  \right)^{-1} \ .
\end{equation}
which is the maximum growth rate achievable in a nutrient condition characterized by $\kappa_n$. Expressing $\kappa_n$ and $\kappa_t$ as function of $\lambda^*$ and $\lambda_{\max}$ we arrive to a simple expression
\begin{equation}
\frac{\lambda(f)}{\lambda^*} = f  \left( 1 - (1-f) \frac{\lambda^*}{\lambda_{\max}} \right)^{-1} \ ,
\label{si.eq:grate_f}
\end{equation}
which defined the growth rate of a population allocating a fraction $f$ of the available resources to the present condition. Figure~\ref{sfig:lambdaTvsf}A shows the dependency of $\lambda/\lambda^*$ on $f$ and $\lambda^*/\lambda_{\max}$ . The function increases with $f$ and, as expected, reaches $1$ when $f=1$. Interestingly, for fast nutrient condition $\lambda^* \approx \lambda_{\max}$, there is a higher range of values of $f$ corresponding to growth close to the maximal possible $\lambda \sim \lambda^*$. 
On the other hand, for slower growth conditions $\lambda^* \ll \lambda_{\max}$, allocating fewer resources to the present condition has a higher impact on growth (see Figure~\ref{sfig:lambdaTvsf}A).

It is important to notice that $\phi_R$ (and $\phi_P$) do depend on the value of $f$. 
In particular, using eq.~\ref{sieq:allocationfluxes} and $\phi_R + \phi_P = \phi_{\max}$, we obtain
\begin{equation}
\phi_R=  \frac{\lambda}{\kappa_t} =  \phi_{\max} \frac{\lambda}{\lambda_{\max}} = \phi_{\max} \frac{f \lambda^*}{\lambda_{\max} - (1-f)\lambda^*} \ ,
\label{si.eq:R_f}
\end{equation}
and, similarly,
\begin{equation}
\phi_P = \phi_{\max} - \phi_R= 
\phi_{\max} \frac{\lambda_{\max} - \lambda^*}{\lambda_{\max}-(1-f)\lambda^*}  =
\phi_{\max} \frac{\lambda_{\max} - \lambda^*}{\lambda_{\max}-(1-f)\lambda^*}  \ .
\end{equation}
Changing $f$ has, therefore, a twofold effect: it selects the fraction of $\phi_P$ allocated to the present condition and, by affecting the value of the growth rate $\lambda$, redistributes the allocation between $\phi_R$ and $\phi_P$ within $\phi_{\max}$. 

We can also write
\begin{equation}
\lambda = \lambda_{\max} \frac{\phi_R}{\phi_{\max}} =
\lambda_{\max} \left( 1- \frac{\phi_P}{\phi_{\max}} \right)  \ .
\end{equation}
By inverting equation~\ref{si.eq:grate_f}, one can derive the dependency of $f$ on the growth rate, obtaining
\begin{equation}
f = \frac{\lambda_{\max}-\lambda^*}{\lambda^*} \frac{\lambda(f)}{\lambda_{\max} - \lambda(f)} \ .
\label{si.eq:f_grate}
\end{equation}
By using $\phi_P = \phi_{\max} (1-\lambda(f)/\lambda_{\max})$ one obtains
\begin{equation}
\phi_{P,\now} =f \phi_P =
\phi_{\max} \frac{\lambda(f)}{\lambda_{\max}}  \frac{\lambda_{\max}-\lambda^*}{\lambda^*}
\ ,
\label{si.eq:phinow_grate}
\end{equation}
and
\begin{equation}
\phi_{P,\notnow} =(1-f) \phi_P =
\phi_{\max} \left( 1 - \frac{\lambda(f)}{\lambda^*} \right)
\ .
\label{si.eq:phinotnow_grate}
\end{equation}
By inverting this expression one obtains
\begin{equation}
\frac{\lambda(f)}{\lambda^*} = 
1- \frac{\phi_{P,\notnow}}{
\phi_{\max} }
\ ,
\end{equation}
which is equivalent to what obtained for the reduction in growth rate due to the expression of a useless protein~\cite{scott2010interdependence}, since $\phi_{P,\notnow}$ can indeed be interpreted as proteins that are useless for the current growth. 

\subsection{Effect of initial proteome allocation on lag-time}
\label{si.sec.physiology.lag}

When conditions change, cells need to adapt their proteome allocation to meet the new environment. Just after the condition changes, a cell will have an initial proteome allocation needed for growth in the present condition equal to $\phi_{P,\new}(0)$. The fraction $\phi_{P,\new}(t)$ will increase over time and (asymptotically) reach a value $\phi_{P,\new}(\infty)$, which corresponds to the growth rate in balanced exponential growth $\lambda(\infty) = \kappa_n \phi_{P,\new}(\infty)$.

We assume that $\phi_{P,\new}(t)$ follows the following dynamics
\begin{equation}    
\frac{d \phi_{P,\new}(t) }{dt } = \frac{1}{\tau}
\phi_{P,\new}(\infty) h\left(
\frac{\phi_{P,\new}(t)}{\phi_{P,\new}(\infty)}
\right)
\left(
1 - \frac{\phi_{P,\new}(t)}{\phi_{P,\new}(\infty)}
\right) \ .
\end{equation}
The last term $(
1 - \phi_{P,\new}(t)/\phi_{P,\new}(\infty))$ ensures that, a long run, the value $\phi_{P,\new}(t)$ converges to $\phi_{P,\new}(\infty)$. The term $
\phi_{P,\new}(\infty) h\left(
\phi_{P,\new}(t)/\phi_{P,\new}(\infty) 
\right)/ \tau$ determines the rate of growth of $\phi_{P,\new}$ as function of $\phi_{P,\new}$ itself. The parameter $\tau$ defines the time-scale associated with physiological adaptation.

A natural assumption is that the rate of change of the proteome sector is limited by the sector itself, as the nutrient flux limits growth. 
This case would correspond to $h(x) = x$, which, for small $\phi_{P,\new}$, implies $d\phi_{P,\new}/dt \propto \phi_{P,\new}$.
This dependence is effectively equivalent to the one proposed in~\cite{erickson2017global}, leading to the same logarithmic dependency on the pre-allocation. 
An alternative to this dependence is the one observed for glycolytic to gluconeogenic shifts, where $h(x) = x^2$~\cite{basan2020universal}.


We further assume that the growth rate $\mu_\new(t)$ is time-dependent as it is limited by the flux of nutrients: $\mu_\new(t) = \kappa_n \phi_{P,\new}(t)$. This time-dependence --- an initial reduced growth rate due to the physiological adaptation to the new environment --- phenomenologically corresponds to a lag time $T$ defined as
\begin{equation}
\int_0^t ds \mu_\new(s) = \lambda (t - T)
\end{equation}
where $\lambda = \mu_{\new}(\infty) = \kappa_n \phi_{P,\new}(\infty)$, which leads to the definition
\begin{equation}
T = \int_0^\infty ds \  \left( 1- \frac{\mu_{\new}(s)}{\lambda} \right) \ .
\end{equation}

We can write
\begin{equation}
T = \int_{\phi_{P,\new}(0)}^{\phi_{P,\new}(\infty)} d \phi_{P,\new} \ \left( \frac{d \phi_{P,\new}(t) }{dt } \right)^{-1} \left( 1- \frac{\phi_{P,\new}(s)}{\phi_{P,\new}(\infty)} \right) = \tau
\int_{\phi_{P,\new}(0)/\phi_{P,\new}(\infty)}^1 \frac{d z}{h(z)} 
\ ,
\end{equation}
and finally obtain
\begin{equation}
T = \tau G\left( \frac{ \phi_{P,\new}(0) }{\phi_{P,\new}(\infty)} \right)    \ ,
\label{si.eq:lag_prealloc}
\end{equation}
where 
\begin{equation}
G(x) = \int_x^1 \frac{dz}{h(z)}  \ .
\end{equation}
We consider a general functional form $h(z) = z^\theta$, which corresponds to
\begin{equation}
G(x) = \frac{x^{1-\theta}-1}{\theta-1} \ ,
\label{si.eq:g_form}
\end{equation}
where the case $\theta=1$ is interpreted by analytical continuation as $G(x) = -\log x$. 

In summary, the lag time is a decreasing function of the ratio between the initial mass fraction $\phi_{P,\new}(0)$ and the fraction corresponding to balanced exponential growth $\phi_{\new}(\infty)$.

\section{Change of condition and growth-lag trade-off}
\label{si.sec.tradeoff}

When a condition changes, cells experience a lag time that depends on the initial allocation $\phi_{P,\new}(0)$. We assume that this initial allocation is equal to the proteome fraction allocated for that condition in the previous one (which is proportional to $\phi_{P,\notnow}$ of the previous condition).
Increasing the value of $\phi_{P,\notnow}$ has therefore two effects: 1) a disadvantageous one, as it decrease the growth rate, because it devotes resources to processes not needed in the condition cells are experiencing (as described in section~\ref{si.sec.physiology.prealloc}), 2) a beneficial one, as it decreases the lag time by increasing the initial allocation $\phi_{P,\new}(0)$ when a new condition arrives (as described in section~\ref{si.sec.physiology.lag}).

Let us consider the case of multiple conditions $a=1,\dots,M$. Cells allocate their proteome sector $P$ to these multiple conditions. Following the experimental evidence in~\cite{mukherjee2024plasticity}, we assume that the intrinsic nutrient qualities are equal ($\kappa_{n,a} = \kappa_{n,b} = \kappa_{n}$) and, as a consequence, the maximal growth rate is condition-independent ($\lambda_a^* = \lambda_b^* = \lambda^*$).

When growing in $a$, cells express a mass fraction of proteins needed to grow in $b$ as part of $\phi_{P,\notnow}$, as defined in section~\ref{si.sec.physiology.prealloc}. More specifically, while growing in $a$, a cell will allocate a proteome fraction $\phi_{a|a}$ to the conditions of $a$ (corresponding to $\phi_{P,\now}$) and a fraction $\phi_{b|a}$ to the condition $b$. We express these proteome fractions as
\begin{equation}
\phi_{b|a} = \phi_{P,a} \left( 
f_a \delta_{a,b} + q_{b|a} (1-f_a) (1-\delta_{a,b}) 
\right) \ ,
\end{equation}
where $\phi_{P,a} = \phi_{\max} - \phi_{R,a}$ is the mass fraction allocated to the sector $P$ when growing in condition $a$, $f_a$ is the fraction of the sector $P$ allocated to the proteins needed for growth in condition $a$, and $q_{b|a}$ is the fraction of $\phi_{P,\notnow}$ allocated to $b$ when in condition $a$.
 Equation~\ref{si.eq:grate_f} can be written under this notation as
\begin{equation}
\frac{\lambda_a(f_a)}{\lambda^*} = f_a  \left( 1 - (1-f_a) \frac{\lambda^*}{\lambda_{\max}} \right)^{-1}
\end{equation}
where 
\begin{equation}
\lambda^* \equiv \lambda(f_a=1) = \phi_{\max}
\left( 
\frac{1}{\kappa_t} + \frac{1}{\kappa_{n}}
\right)^{-1}
\end{equation}

At the same time, when conditions switch from $a$ to $b$, cells will experience a lag $T_{b|a}$ whose duration is defined by equation~\ref{si.eq:lag_prealloc} that can be written as
\begin{equation}
T_{b|a} = \tau G\left( \frac{\phi_{b|a}}{\phi_{b|b}} \right) \ ,
\label{si.eq:lagratio}
\end{equation}
where $\phi_{b|b} = f_b \phi_{P,b}$ is the proteome sector allocated to growth in $b$ when the condition equals to $b$.

Using equation~\ref{si.eq:phinotnow_grate}, we can write
\begin{equation}
\phi_{b|a} = q_{b|a} \phi_{\max} \left( 1 - \frac{\lambda_a(f_a)}{\lambda_a^*} \right) \ ,
\end{equation}
and, using equation~\ref{si.eq:phinow_grate}, we have that
\begin{equation}
\phi_{b|b} = 
\phi_{\max} \frac{\lambda(f_b)}{\lambda_{\max}}  \frac{\lambda_{\max}-\lambda^*}{\lambda^*}
\ .
\end{equation}
Putting these two equations together, we obtain
\begin{equation}
\frac{\phi_{b|a}}{\phi_{b|b}} =
 q_{b|a} \left(
1 - \frac{\lambda(f_a)}{\lambda^*}
\right)
\frac{\lambda^*}{\lambda(f_b)} \left(
1-\frac{\lambda^*}{\lambda_{\max}}
\right)^{-1}
\ .
\label{si.eq:ratio_general}
\end{equation}
The lag-time is a decreasing function of this ratio (as defined in equation~\ref{si.eq:lagratio}).
This relation implies that the lag-time increases with the growth rate $\lambda(f_a)$, determining a trade-off between growth and lag. At the same time, the lag time increases with the growth of the post-shift condition $\lambda(f_b)$.

One can also express this ratio as a function of the allocation parameters $f$, by using eq.~\ref{si.eq:grate_f} from which one obtains
\begin{equation}
\frac{\phi_{b|a}}{\phi_{b|b}} =
 q_{b|a} \left(
1 - \frac{f_a}{1-(1-f_a)\lambda^*/\lambda_{\max}}
\right)
\frac{1-(1-f_b)\lambda^*/\lambda_{\max}}{f_b} \left(
1-\frac{\lambda^*}{\lambda_{\max}}
\right)^{-1}
\ .
\end{equation}

These expressions simplify if the two conditions have the same allocation parameters $f_a = f_b = f$ (and, consequently, the same growth rates). In this case, one obtains
\begin{equation}
\frac{\phi_{b|a}}{\phi_{b|b}} =
 q_{b|a} \left(
\frac{\lambda^*}{\lambda(f)} - 1
\right)
 \left(
1-\frac{\lambda^*}{\lambda_{\max}}
\right)^{-1}
\ .
\label{sieq:ratio}
\end{equation}

Fig.~\ref{sfig:lambdaTvsf} shows the dependency of the inverse lag-time $1/T_{b|a}$ as function of the allocation parameter $f$ for $\theta = 1$ and $\theta = 2$ and different values of $\lambda^*/\lambda_{\max}$, which is defined as function of the ratio $\phi_{b|a}/\phi_{b|b}$ using eq.~\ref{si.eq:lagratio}.

\section{Growth in serial dilution}
\label{secsi.growthserial}

We consider a population with abundance $n(t)$ that grows on a given resource with a maximum growth rate $\lambda$ after a lag-time $T$. When the population grows, the abundance changes according to
\begin{equation}
\frac{dn}{dt} = \lambda n u(c/K)    \ \ \  \text{for $t > T$}
\end{equation}
and the concentration according to
\begin{equation}
\frac{dc}{dt} = -\lambda \frac{n}{Y} u(c/K) \ \ \ \text{for $t > T$}  \ .
\end{equation}
The function $u(\cdot)$ encodes the functional response (how the growth of the population depends on the concentration of the resource in the environment). We define $\lim_{z \to \infty }u(z) = 1$ (so then $\lambda$ is the growth rate at large concentrations) and $u(1)=1/2$, so that $K$ can be interpreted as the concentration when the growth rate is equal to half the maximum. A common assumption is Monod's form $u(z) = z/(1+z)$.

From the equation above, it is easy to see that the abundance is strictly related to the resource concentration in the environment via
\begin{equation}
    n(t) = n(0) + Y \left( c_0 - c(t) \right) \ ,
\end{equation}
where $c_0$ is the initial resource concentration.

In the following, we assume that $c_0 \gg K$. Under this condition we can approximate $u(z) \approx 1$ in the range where $c(t) \gg K$ to obtain
\begin{equation}
    n(t) = n(0) e^{\lambda \Phi(t-T) } \ ,
\end{equation}
where $\Phi(z) = z$ if $z>0$ and $0$ otherwise. The corresponding value of the resource concentration will be equal to
\begin{equation}
    c(t) = c_0 - \left( e^{\lambda \Phi(t-T) } - 1 \right) \ .
\end{equation}
The growth of $n$ will slow down and stop when $c(t) \approx K$. In the limit $K \ll c_0$ this will occur at a the depletion time $t_s$ equal to
\begin{equation}
t_s = T + \log \left( 1 + \frac{Y c_0}{n(0)} \right) \ .
\end{equation}

We consider a serial-dilution (or boom-and-burst) environment, when, following a cycle of growth, a population is reduced (``diluted'') of a factor $D$. If we track the initial population $n^d(0)$ at the cycle $d$ and the final one $n^d(t_f)$, we will have that, assuming that all the resources are depleted ($t_f > t_s^d$)
\begin{equation}
D n^{d+1}(0) = n^{d}(t_f)= n^d(0) + Y c_0 \ .
\label{sieq.newabd}
\end{equation}
Tracked over cycles, the population converges to a stationary value $n^*(0)$ equal to
\begin{equation}
 n^{*}(0) = \frac{ Y  c_0 }{D-1} \ .
\end{equation}
The corresponding stationary consumption time equals
\begin{equation}
    t_s = T + \frac{1}{\lambda} \log D \ .
\end{equation}

\section{Invasion analysis in deterministic boom-and-bust environments}

In order to establish the outcome of evolutionary dynamics, we perform invasion analysis. In this setting, we assume that a given strain $i$ reached stationarity and a strain $j$ is introduced at infinitesimal abundance at the beginning of the cycle. Strain $j$ is able to invade population $i$ if, at the next cycle, its initial population grows. The population of the invader grows exponentially with growth rate $\lambda_j$ and lag-time $T_j$
\begin{equation}
n_j(t) = n(0) \exp\left( \lambda_j \Phi(t-T_j) \right) \ .    
\end{equation}
The growth stops when resources are fully consumed. This occurs at time $t^*_{s,i}$ which is set by the resident population $i$, since $j$ is in low abundance and does not affect significantly the resource concentration
\begin{equation}
t_{s,i} = T_i  + \frac{1}{\lambda_i} \log D \ .    
\end{equation}

Following a dilution of a factor $D$, the abundance at the next cycle of the invader equals
\begin{equation}
n_j^{d+1}(0) = \frac{ n^d(0) \exp\left( \lambda_j \Phi(t_{s,i}-T_j) \right) }{D} \ .   
\label{si.eq:invaderpop}
\end{equation}
The invasion criterion $n^{d+1}_j(0) > n^d_j(0)$ reads
\begin{equation}
    \lambda_j \Phi(t_{s,i}-T_j) > \log D \ .
\end{equation}

On the other hand, by definition of the consumption time, we have
\begin{equation}
t_{s,j} = T_j + \frac{\log D}{ \lambda_j} \ ,
\end{equation}
and therefore, expressing $\log D$ as function of $\lambda_j$, $T_j$ and $T_j$, we obtain that the strain $j$ is able to invade if
\begin{equation}
    t_{s,j} - T_j <
     \Phi(t_{s,i}-T_j) 
\label{si.eq:invdet}
    \ .
\end{equation}

Note that $t_{s,i} - T_j > 0$ if
\begin{equation}
\frac{\log D}{\lambda_j} > T_j - T_i
    \ .
\end{equation}
If we assume that strain $i$ and $j$ are not too dissimilar, we expect the difference of their lag-time to be small (compared to the time spent growing). In that case, the condition for invasion simplifies to
\begin{equation}
    t_{s,j}  < t_{s,i}  
    \ .
\end{equation}
This result implies that strains with lower resource depletion time are evolutionary advantaged.

\section{Emergence of an evolutionary stable state in deterministic boom-and-bust environments}

If strains differ because of their allocation strategy $f_i$, we expect evolution to drive the systems toward values of the allocation minimizing the depletion time.
We consider here the case of two conditions, with equal allocation values $f$. In this scenario, the growth rate $\lambda$ is a function of $f$ expressed in equation~\ref{si.eq:grate_f}. Similarly, the lag-time is a function of the allocation value $f$ (or equivalently of the growth rate) as expressed in section~\ref{si.sec.tradeoff}.

Under this setting, the depletion time of strain $i$ can be written as
\begin{equation}
t_{s,i} = T(\lambda_i) + \frac{\log D}{ \lambda_i} \ ,
\end{equation}
where $T(\lambda_i)$ expresses the lag-time as a function of the growth rate $\lambda_i$ (which can be equivalently thought of as a function of strain $i$-'s allocation $f$).
Using eq.~\ref{sieq:ratio}, we have the dependency
\begin{equation}
T(\lambda) = \tau  G \left( \frac{\lambda^*-\lambda}{\lambda} \left(1-\frac{\lambda^*}{\lambda_{\max}}\right)^{-1} \right) \ ,
\label{eq: lag-time-dep}
\end{equation}
where we considered $q=1$.

From this expression, we can derive the value of $\lambda$ corresponding to the minimum of the depletion time
\begin{equation}
t^{\min} = \min_{\lambda \in [0,\lambda^*]} \left( T(\lambda) + \frac{\log D}{\lambda} \right) \ .
\end{equation}
The value of the growth rate minimizing the depletion time
\begin{equation}
\lambda^{\ess} = \text{argmin}_{\lambda \in [0,\lambda^*]} \left( T(\lambda) + \frac{\log D}{\lambda} \right) 
\end{equation}
is the evolutionary stable state of the system. The invasion condition defined in eq.~\ref{si.eq:invdet} implies that $\lambda^{\ess}$ is an uninvasible evolutionary stable state. We expect evolutionary trajectories to converge to $\lambda^{\ess}$  over large evolutionary times.

By assuming the dependency of the lag time given in eq.~\ref{eq: lag-time-dep} with $G(z) = (z^{1-\theta}-1)/(\theta-1)$ as described in section~\ref{si.sec.physiology.lag}, we obtain
\begin{equation}
\frac{\lambda^*}{\lambda^{\ess}} = 1 + \left(
\frac{\tau \lambda^*}{\log D}
\right)^{1/\theta}\left(
1-\frac{\lambda^*}{\lambda_{\max}}
\right)^{1/\theta-1}  \ .
\label{si.eq.lambdaess}
\end{equation}

\section{Invasion analysis in stochastic boom-and-bust environments}
\label{secsi.invstoch}

We consider a stochastic boom-and-bust environment. Stochasticity is determined by three factors: randomness in the resource supply, randomness in the dilution factor, and randomness in the conditions under which cells are growing. All three of these factors are therefore cycle-dependent. 

As a consequence of stochasticity, the population abundance of a strain growing in isolation at the beginning of every cycle will be cycle-dependent. Assuming that resources are always depleted, we can still use equation~\ref{sieq.newabd}, which is generalized to
\begin{equation}
n^{d+1}(0) = \frac{n^d(0) + Y c_0^d}{D^d} \ ,
\end{equation}
where the resource supply $c_0^d$ and the dilution factor $D^d$ both depend on cycle $d$. 

The depletion time will be equal to
\begin{equation}
t_s^d = T_{E^d|E^{d-1}} + \frac{1}{\lambda_{E^d}}\log \frac{n^d(0) + Y c_0^d}{n^d(0)}\ ,
\label{eqsi.deptime}
\end{equation}
where $E^d$ is the condition in cycle $d$ and $E^{d-1}$ the condition in the previous cycle. The second term represents the time needed to grow from the initial population $n^d(0)$ to a final population $n^d(0) + Y c_0^d$ with a condition-dependent  growth rate $\lambda_{E^d}$.

We consider $c_0^d$ and $D^d$ drawn randomly and independently from two specified probability distributions.
We assume that the conditions occur in a random sequence. The statistical structure of the environment is encoded in the transition matrix $\mathbf{w}$ whose elements $w_{a|b}$ specify the probability that condition $a$ follows condition $b$. By normalization $\sum_a w_{a|b} = 1$. We assume that a new condition is never repeated ($w_{a|a}=0$). Note that this assumption does not reduce the generality of the framework. In our framework, repeating a condition for two cycles is equivalent to considering a single cycle where resources are supplied twice and dilution occurs twice.
The transition probability corresponds to a stationary distribution $p_c$ obeying $\sum_{b} w_{a|b} p_b = p_a$, which corresponds to the probability that condition $a$ occurs.

By averaging over cycles eq.~\ref{eqsi.deptime} we can calculate the average depletion time.
We can define the expected depletion time conditioned on growth in $a$
\begin{equation}
\mathbb{E}_a\left( t_s \right) =
\sum_{b} w^{\leftarrow}_{b|a} T_{a|b} + 
\frac{1}{\lambda_a} \mathbb{E}_a\left( \log \frac{n^d(0) + Y c_0^d}{n^d(0)} \right) =
\frac{1}{p_a} \sum_{b} w_{a|b} p_b T_{a|b} + 
\frac{1}{\lambda_a} \mathbb{E}_a\left( \log \frac{n^d(0) + Y c_0^d}{n^d(0)} \right)
\label{sieq.expectedtcond}
\end{equation}
where $w^{\leftarrow}_{b|a}$ is the probability  that condition $a$ was preceded by condition $b$, which can be expressed using Bayes' theorem as 
$w^{\leftarrow}_{b|a} p_a = w_{a|b} p_b$.

The expected depletion time (averaged over conditions) reads
\begin{equation}
\mathbb{E}[t_s] = \sum_a p_a \mathbb{E}_a\left[ t_s \right] = \sum_a p_a \sum_{b} w_{a|b} p_b T_{a|b} + \sum_a \frac{p_a}{\lambda_a} \mathbb{E}_a\left[ \log \frac{n^d(0) + Y c_0^d}{n^d(0)} \right] \ .
\end{equation}

If the dilution factor $D$ and the resource supply $c_0$ are resource independent, we have that
\begin{equation}
 \mathbb{E}_a\left[ \log \frac{n^d(0) + Y c_0^d}{n^d(0)} \right] =  \mathbb{E}\left[ \log \frac{n^d(0) + Y c_0^d}{n^d(0)} \right] =
  \mathbb{E}\left[ \log D^d \frac{n^{d+1}(0)}{n^d(0)} \right] = 
    \mathbb{E}\left[ \log D^d  \right] 
 \ ,
\end{equation}
where we used the fact that $\mathbb{E}\left[ \log n^{d+1}(0) - \log n^d(0) \right] = 0$ at stationarity. Putting these expression together we obtain that
\begin{equation}
\mathbb{E}[t_s] = \sum_a \sum_{b} w_{a|b} T_{a|b} p_b + \sum_a \frac{p_a}{\lambda_a} \mathbb{E}\left[ \log D \right] \ ,
\label{si.eq.expectedtimedep}
\end{equation}
whenever the dilution factor and resource supply are condition-independent.

In the symmetric case $w_{a|b}=1/(M-1)$ (which corresponds to $q_{a|b}=1/(M-1)$) and $\lambda_a =\lambda$, eq.~\ref{si.eq.expectedtimedep} simplifies to
\begin{equation}
\mathbb{E}[t_s] = \tau G\left( \frac{1}{M-1} \frac{\frac{\lambda^*}{\lambda}-1}{1-\frac{\lambda^*}{\lambda_{\max}}} \right) +  \frac{\mathbb{E}\left[ \log D \right] }{\lambda} \ .
\end{equation}
The minimum of the depletion time is achieved for
\begin{equation}
\frac{\lambda^*}{\lambda^{\ess}} =
 1 + \left(
\frac{\tau \lambda^*}{\mathbb{E}[\log D]}
\right)^{1/\theta}\left( (M-1)\left(
1-\frac{\lambda^*}{\lambda_{\max}}
\right)\right)^{1/\theta-1} 
\ .
\label{sieq.lambdaessstochsym}
\end{equation}


\subsection{Invasibility condition}
\label{secsi.invasionfull}

Let us consider an invader (or mutant) strain with allocation values $\underline{f}'$ and $\mathbf{q}'$ and initial population $n_i^{d=0}(0)$. The allocations $\underline{f}'$ uniquely determine a corresponding set of growth rates $\underline{\lambda}'$, where the growth rate $\lambda_a' = \lambda(f_a')$ as given by eq.~\ref{si.eq:grate_f}. The resident strain has growth rates $\underline{\lambda}$ and preallocations $\mathbf{q}$.

The traits of the resident strain $\underline{\lambda}$ and $\mathbf{q}$ determine the depletion time $t_s^d(\underline{\lambda},\mathbf{q})$ according to~\ref{eqsi.deptime}
\begin{equation}
t_s^d(\underline{\lambda},\mathbf{q}) = T_{E^d|E^{d-1}}(\underline{\lambda},\mathbf{q}) + \frac{\log D^d + \log n^{d+1}(0) -\log n^{d}(0) }{\lambda_{E^d}}   \ ,
\end{equation}
where $n^d(0)$ is the abundance of the resident strain at the beginning of cycle $d$ and $E^d$ is the condition at cycle $d$.
The invader strain will start growing after its lag-time and stop at the depletion time, which gives
\begin{eqnarray}
\log \frac{n_i^{d+1}(0)}{n_i^{d}(0)} && = 
\lambda'_{E^d} \Phi\left( 
t_s^d(\underline{\lambda},\mathbf{q}) -
T_{E^d|E^{d-1}}(\underline{\lambda}',\mathbf{q}')
\right) - \log D^d = \\ \nonumber
&& = \lambda'_{E^d} \Phi\left( 
t_s^d(\underline{\lambda},\mathbf{q}) -
T_{E^d|E^{d-1}}(\underline{\lambda}',\mathbf{q}')
\right) - 
\lambda'_{E^d} \left( 
t_s^d(\underline{\lambda}',\mathbf{q}') -
T_{E^d|E^{d-1}}(\underline{\lambda}',\mathbf{q}')
\right) - \log \frac{n^{d+1}(0)}{n^d(0)}
\ .
\label{sieq.growthinvader}
\end{eqnarray}
A strain appearing at cycle $d=0$, will invade if $n^{d}(0) > n^{d=0}(0)$ for large $d$. Using
\begin{equation}
0 <  \log \frac{n_i^{d}(0)}{n_i^{d=0}(0)} = \sum_{g=0}^{d-1} \log \frac{n_i^{g}(0)}{n_i^{g-1}(0)} \to d \mathbb{E}\left[\log \frac{n_i^{g}(0)}{n_i^{g-1}(0)} \right] \  ,
\end{equation}
together with~\ref{sieq.growthinvader} --- also assuming that the two strains are not too dissimilar (which allows to consider $\Phi(x) = x$) --- we obtain the invasion criterion
\begin{equation}
  \mathbb{E}\left[
\lambda' \left( t_s(\underline{\lambda},\mathbf{q}) -
t_s(\underline{\lambda}',\mathbf{q}') \right)
\right] > 0 \  ,
\label{sieq.invasioncrit}
\end{equation}
where we used the stationarity condition for the resident population
\begin{equation}
  \mathbb{E}\left[
\log \frac{n^{d+1}(0)}{n^d(0)}
\right] = 0 \  .
\end{equation}

Eq.~\ref{sieq.invasioncrit} can be explicitly written in terms of the conditional averages over conditions
\begin{equation}
  \mathbb{E}\left[
\lambda' \left( t_s(\underline{\lambda},\mathbf{q}) -
t_s(\underline{\lambda}',\mathbf{q}') \right)
\right] = \sum_a p_a \lambda'_a
  \left(\mathbb{E}_a\left[t_s(\underline{\lambda},\mathbf{q})\right] -
\mathbb{E}_a\left[t_s(\underline{\lambda}',\mathbf{q}')
\right] \right) > 0
\  .
\label{sieq.invasioncrit_cond}
\end{equation}

\subsection{Evolutionary-stable state for symmetric conditions}
\label{secsi.invationsym}

In this section, we consider the scenario where conditions are equivalent. More precisely, we consider the ansatz where the evolutionary stable state $\underline{f}^{\ess}$ (and the corresponding values of the growth rate) take the same value irrespective of the condition, i.e. $f_a^{\ess} = f^{\ess,eq}$ , for all $a=1,\dots,M$. Under this choice, also the growth rates become condition independent $\lambda(f_a^{\ess}) = \lambda(f^{\ess,eq})$, and eq.~\ref{sieq.invasioncrit_cond} simplifies to a condition on the expected depletion times
\begin{equation}
  \mathbb{E}\left[ t_s(\underline{\lambda}',\mathbf{q}')  \right] <
  \mathbb{E}\left[ t_s(\underline{\lambda},\mathbf{q}) \right] \  .
\label{sieq.invasioncrit}
\end{equation}
A strain invades another if its expected depletion time is lower than than of the resident.

\section{A primer on adaptive dynamics}
\label{secsi.adaptiveprimer}

In this section we consider a general scenario of eco-evolutionary dynamics which we study under the framework of adaptive dynamics~\cite{metz1995adaptive,diekmann2002beginners}. This approach assumes as separation between ecological and evolutionary dynamics, aiming at describing the latter. Mutations are rare and strongly selected: mutation always go to fixation or extinction before a new mutation arises.

Let us consider an  population where individuals are characterized by trait $\underline{x}$. Let $P(\underline{x},t)$ be the probability that a randomly chosen individual in the population has trait  $\underline{x}$ at time $t$. We expect this fraction to evolve accordingly to the quasi-species model~\cite{bull2005quasispecies}
\begin{equation}
  \frac{\partial P(\underline{x},t)}{\partial t} = 
  \int d \underline{x}' \ 
  \left(
  W(\underline{x}|\underline{x}') P(\underline{x}',t) -
    W(\underline{x}'|\underline{x}) P(\underline{x},t)
\right) \ ,
\label{sieq.mastereq}
\end{equation}
where $W(\underline{x}|\underline{x}')$ captures the effect of mutations and selection. In particular, we decompose
\begin{equation}
  W(\underline{x}'|\underline{x}) =
  M(\underline{x}'|\underline{x})
  D(\underline{x}'|\underline{x})
  \ ,
\end{equation}
where $M(\underline{x}'|\underline{x})$ is the rate at which an individual with trait $\underline{x}$ reproduces in a mutated offspring with trait $\underline{x}'$ and $D(\underline{x}'|\underline{x})$ captures the effect of selection. 

We can write the following equation for average trait $\langle \underline{x} \rangle$ in the population
\begin{equation}
  \frac{d \langle x \rangle}{d t} = 
 \int d \underline{x} \ \int d \underline{x}' \ \underline{x} \ 
  \left(
  W(\underline{x}|\underline{x}') P(\underline{x}',t) -
    W(\underline{x}'|\underline{x}) P(\underline{x},t)
\right) = 
 \int d \underline{x} \ \int d \underline{x}' \ (\underline{x}'-\underline{x}) \ 
    W(\underline{x}'|\underline{x}) P(\underline{x},t)
\ .
\label{sieq.mastereq_mean}
\end{equation}
Adaptive dynamics assumes that the population is monomorphic, i.e. $P(x,t) = \delta( x - \langle x \rangle_t )$. From eq.~\ref{sieq.mastereq} we can write
\begin{equation}
  \frac{d \langle x \rangle}{d t} = 
  \int d \underline{x}' \ (\underline{x}'-\langle\underline{x}\rangle) \ 
    W(\underline{x}'|\langle\underline{x}\rangle)
\ .
\label{sieq.mastereq_meancanonical}
\end{equation}
In a monomorphic population with trait $\underline{x}$ we can write the mutation rate as
\begin{equation}
  M(\underline{x}'|\underline{x}) = 
  b(\underline{x}) N^*(\underline{x}) U(\langle\underline{x}\rangle) K(\underline{x}'-\underline{x})
  \ ,
\end{equation}
where we consider that the resident population has birth rate $b(\underline{x})$, population size $N^*(\underline{x})$, mutation rate $U$. We also assume that when a mutation occurs, the new trait will be distributed according to the mutation kernel $K(\underline{x}'-\underline{x})$ (which we imagine to be centered around $\underline{0}$).
The selection effect will be captured by $D(\underline{x}'|\underline{x})$, which is interpreted as the fixation probability of a mutant with trait $\underline{x}'$ emerging in a resident population with trait $\underline{x}$. In a large population, the fixation probability~\cite{kimura1962probability} can be written as
\begin{equation}
D(\underline{x}'|\underline{x}) = 
\max \left\{ 0, 1-\frac{d(\underline{x}'|\underline{x})}{b(\underline{x}'|\underline{x})} \right\}
  \ ,
\end{equation}
where $b(\underline{x}'|\underline{x})$ and $d(\underline{x}'|\underline{x})$ are birth and death rate of a mutant with trait $\underline{x}'$ when the resident has trait $\underline{x}$. We define $s_{\underline{x}}(\underline{x}')=b(\underline{x}'|\underline{x})-d(\underline{x}'|\underline{x})$ as the \textit{invasion fitness}, which corresponds to the growth rate when rare.

By considering these terms together in equation~\ref{sieq.mastereq_meancanonical}, we obtain
\begin{equation}
  \frac{d \langle \underline{x} \rangle}{d t} = 
  N^*(\langle\underline{x}\rangle) U
  \int d \underline{x}' \  \ 
     \frac{b(\langle\underline{x})\rangle}{b(\underline{x}'|\langle\underline{x}\rangle)}  K(\underline{x}'-\langle\underline{x}\rangle) (\underline{x}'-\langle\underline{x}\rangle)
     \max \left\{ 0, s_{\langle\underline{x}\rangle}(\underline{x}') \right\}
\ .
\end{equation}
We assume that the mutation kernel is peaked around zero (mutations are small), and therefore we can expand around $\underline{x}' \approx \langle \underline{x} \rangle$. We obtain for the $\alpha$-th trait
\begin{equation}
  \frac{d \langle x_\alpha \rangle}{d t} = 
  N^*(\langle\underline{x}\rangle) U(\langle\underline{x}\rangle) 
  \int d \underline{x}' \  \  K(\underline{x}'-\langle\underline{x}\rangle) (x'_\alpha-\langle x_\alpha \rangle)
     \max \left\{ 0, \sum_\beta (x_\beta'-\langle x_\beta \rangle) \frac{\partial s_{\langle\underline{x}\rangle}(\underline{y}) }{\partial y_\beta} \biggr|_{\underline{y} = \langle\underline{x}\rangle} \right\}
\ ,
\end{equation}
where we used $b(\underline{x}'|\langle\underline{x}\rangle) \approx b(\langle\underline{x}\rangle)$ and expanded to the first order $s_{\langle\underline{x}\rangle}(\underline{x}')$ around $\underline{x}' = \langle\underline{x}\rangle$,
using the fact that the resident population is at equilibrium (and a neutral mutation has zero invasion fitness)
$s_{\langle\underline{x}\rangle}(\langle\underline{x}\rangle)=0$. In full generality, we can consider the case where mutations are independent across the traits, i.e. $K(\underline{x}'-\underline{x})=\prod_\beta K_\beta(x'_\beta -x_\beta)$
\begin{equation}
  \frac{d \langle x_\alpha \rangle}{d t} = 
  N^*(\langle\underline{x}\rangle) U(\langle\underline{x}\rangle) 
\prod_\alpha  \int d z_\alpha \  \  K_\alpha(z_\alpha) \ z_\alpha
     \max \left\{ 0, \sum_\beta z_\beta \frac{\partial s_{\langle\underline{x}\rangle}(\underline{y}) }{\partial y_\beta} \biggr|_{\underline{y} = \langle\underline{x}\rangle} \right\} = 
\ ,
\end{equation}
where we introduced $z_\beta = x_\beta'-\langle x_\beta \rangle$. Assuming a symmetric mutation kernel $K_\beta(z)=K_\beta(-z)$ we finally obtain the canonical equation of adaptive dynamics
\begin{equation}
  \frac{d \langle x_\alpha \rangle}{d t} = 
 \frac{1}{2} N^*(\langle\underline{x}\rangle) U(\langle\underline{x}\rangle)  \sigma_\alpha^2
 \frac{\partial s_{\langle\underline{x}\rangle}(\underline{y}) }{\partial y_\alpha} \biggr|_{\underline{y} = \langle\underline{x}\rangle}   
\ ,
\label{sieq.canonicaladapt}
\end{equation}
where $\int dz \ K_\beta(z) z^2 = \sigma^2_\beta$ is the variance of the mutation Kernel. The factor $1/2$ emerges from the $\max \{ \cdot \}$ in the integral and the symmetry of the mutation Kernel.

Equation~\ref{sieq.canonicaladapt} defines a non-linear dynamical system for the population average traits $\langle \underline{x} \rangle$.
The term $\frac{\partial s_{\langle\underline{x}\rangle}(\underline{y}) }{\partial y_\alpha} \biggr|_{\underline{y} = \langle\underline{x}\rangle}$ is known as the \textit{selection gradient} and determines the flow of evolutionary dynamics.

Evolutionary equilibria, known as \textit{singular strategies} $\underline{x}^*$, are found where the selection gradient is zero:
\begin{equation}
\frac{\partial s_{\underline{x}^*}(\underline{y}) }{\partial y_\alpha} \biggr|_{\underline{y} = \underline{x}^*} = 0 \ ,
\end{equation}
for all $\alpha$.
This means that at a singular strategy, the fitness landscape is locally flat. Singular strategies corresponds to the fixed points of eq.~\ref{sieq.canonicaladapt}. 
A singular strategy $ \underline{x}^*$ is an \textit{evolutionary stable state} (ESS) if it
cannot be invaded by any nearby mutant. This is true if the curvature of the invasion fitness is negative, i.e. if the matrix
    \begin{equation}
    \frac{\partial^2  s_{\underline{x}^*}(\underline{y})}{\partial  y_\alpha \partial  y_\beta } \biggr|_{\underline{y} = \underline{x}^*} < 0 \ .
    \end{equation}
A singular strategy is \textit{convergently stable} if its attractive, which occurs if the eigenvalues of the matrix
    \begin{equation}
  \left( \frac{\partial}{\partial x_\beta} \frac{\partial s_{\underline{x}}(\underline{y})}{\partial  y_\alpha  } \right)\biggr|_{\underline{y} = \underline{x} = \underline{x}^*} \ ,
    \end{equation}
have all negative real parts. Note that the sign of the eigenvalues of this matrix corresponds to that one of the Jacobian of the dynamics defined in eq.~\ref{sieq.canonicaladapt} evaluated in the singular (fixed) point.


\section{Adaptive dynamics in stochastic boom-and-bust environments}
\label{secsi.adaptiveboom}


Using equation~\ref{sieq.invasioncrit_cond} we can define the expected growth rate
$s_{\underline{\lambda},\mathbf{q}}\left( \underline{\lambda}',\mathbf{q}' \right)$ of an invader with growth-rates $\underline{\lambda}'$ and pre-allocation $\mathbf{q}'$ when the resident population has growth-rate
$\underline{\lambda}$ and pre-allocation $\mathbf{q}$ which in our setting reads
\begin{eqnarray}
s_{\underline{\lambda},\mathbf{q}}\left( \underline{\lambda}',\mathbf{q}' \right) && :=
  \mathbb{E} \left[ \log \frac{n_i^{d+1}(0)}{n_i^{d}(0)} \right] =
  \sum_a p_a 
\lambda'_a  \left(  \mathbb{E}_a \left[ t_{s}\left( \underline{\lambda}',\mathbf{q}' \right) \right] - \mathbb{E}_a \left[ t_{s}\left( \underline{\lambda},\mathbf{q} \right)  \right]
  \right) \ .
\end{eqnarray}

As we discussed above, under the assumption that mutations are rare and have small effects, the trajectory of the population average value of a trait under selection $\langle x \rangle$ can be described by the canonical equation of adaptive dynamics
\begin{equation}
\frac{\partial \langle x \rangle}{\partial t} \propto \frac{\partial}{\partial x'} s_{\langle x \rangle}(x') \biggr|_{x'=\langle x \rangle}   
\label{eqsi.adaptivegeneral}
\ ,
\end{equation}
where the proportionality constant sets the time-scale of evolution and depends on the mutation rate, the mutation kernel, the population size, and other demographic parameters.

In our case, this dynamics translates into the following $M$ equations for $\underline\lambda$
\begin{eqnarray}
  \frac{\partial \langle \lambda_c \rangle}{\partial t}  \propto -
    \sum_a p_a 
\langle \lambda_a \rangle \frac{\partial \mathbb{E}_a \left[ t_{s}\left(  \underline{\lambda}' , \langle\mathbf{q}\rangle \right)\right] }{\partial  \lambda_c' }\biggr|_{\lambda_c'=\langle \lambda_c\rangle} \ ,
\label{eqsi.adaptivelambda}
\end{eqnarray}
one for each condition $c$, and in the following $M(M-1)$ equations for $\bf{q}$
\begin{eqnarray}
  \frac{\partial \langle q_{d|c} \rangle}{\partial t}  \propto -
    \left(\sum_a p_a \langle \lambda_a \rangle \frac{\partial \mathbb{E}_a \left[ t_{s}\left( \langle \underline{\lambda} \rangle, \mathbf{q}' \right) \right]}{\partial q_{d|c}'}\biggr|_{q_{d|c}'=\langle q_{d|c}\rangle} + Z_c \right) \ ,
\label{eqsi.adaptiveq}
\end{eqnarray}
where the term
\begin{equation}
    Z_c = - \sum_d \sum_a p_a 
\langle \lambda_a \rangle \frac{\partial \mathbb{E}_a \left[ t_{s}\left( \langle \underline{\lambda} \rangle, \mathbf{q}' \right) \right]}{\partial q_{d|c}' }\biggr|_{q_{d|c}'=\langle q_{d|c}\rangle} \ ,
\end{equation}
guarantees the normalization $\sum_{a} q_{a|b} = 1$.

Equations~\ref{eqsi.adaptivelambda} and~\ref{eqsi.adaptiveq} define a non-linear dynamical system. The stable fixed point(s) of this dynamical system are the evolutionary stable states. It is important to notice in this context the these equation do not in general admit the existence of a Lyapunov function (i.e., a function which is always minimized by the dynamics).
The evolutionary stable states are the solutions of
\begin{eqnarray}
 0 =
    \sum_a p_a 
 \lambda^{\ess}_a \frac{\partial \mathbb{E}_a \left[ t_{s}\left( \underline{\lambda}' , \mathbf{q}^{\ess} \right) \right] }{\partial  \lambda_c'}\biggr|_{\lambda_c'=\lambda_c^{\ess}} \ 
\label{sieq.lambdaess}
\end{eqnarray}
and
\begin{eqnarray}
Z_c = -
 \sum_a p_a \lambda^{\ess}_a \frac{\partial \mathbb{E}_a \left[ t_{s}\left( \underline{\lambda}^{\ess} , \mathbf{q}' \right) \right]}{\partial q_{d|c}' }\biggr|_{q_{d|c}' = q_{d|c}^{\ess}} \ .
\end{eqnarray}

Using equation~\ref{sieq.expectedtcond} together with equations~\ref{si.eq:lagratio} and~\ref{si.eq:ratio_general}, we obtain
\begin{equation}
 \mathbb{E}_a \left[ t_{s}\left( \underline{\lambda}^{\ess} , \mathbf{q}^{\ess} \right) \right] = 
 \frac{\tau}{p_a} \sum_b w_{a|b} p_b G\left( q_{a|b} \frac{\lambda^*-\lambda_b}{\lambda_a} \frac{\lambda_{\max}}{\lambda_{\max} - \lambda^*}\right) + \frac{\log \bar{D}_a}{\lambda_a} \ ,
 \label{sieq.expecteddepl_condexpl}
\end{equation}
where
\begin{equation}
 \log \bar{D}_a := \mathbb{E}_a\left[ \log \frac{n^d(0) + Y c_0^d}{n^d(0)} \right] = \mathbb{E}_a\left[ \log D \right] +
 \mathbb{E}_a\left[ \log n^{d+1}(0) - \log n^{d}(0)  \right] \ .
\end{equation}
note that the last term $ \mathbb{E}_a\left[ \log n^{d+1}(0) - \log n^{d}(0)  \right] = 0$ unless resource supply is condition dependent.

Equation~\ref{eqsi.adaptiveq} becomes
\begin{eqnarray}
   Z_c && =
 \frac{w_{d|c}\lambda_d}{q_{d|c}} p_c  \left( q_{d|c} \frac{\lambda^*-\lambda_c}{\lambda_d} \frac{\lambda_{\max}}{\lambda_{\max} - \lambda^*}\right)^{1-\theta}  \ ,
 \label{eqsi.Zc_ono}
\end{eqnarray}
which can be solved to write
\begin{eqnarray}
q_{d|c} = \frac{\lambda_d}{Z_c} \left( w_{d|c} p_c \right)^{1/\theta}  \left( (\lambda^*-\lambda_c) \frac{\lambda_{\max}}{\lambda_{\max} - \lambda^*} \right)^{(1-\theta)/\theta}  \ .
\end{eqnarray}
Since $Z_c$ is defined by normalization
\begin{eqnarray}
Z_c = \sum_d \lambda_d \left( w_{d|c} p_c \right)^{1/\theta}  \left( (\lambda^*-\lambda_c) \frac{\lambda_{\max}}{\lambda_{\max} - \lambda^*} \right)^{(1-\theta)/\theta} \ ,
\end{eqnarray}
from which we obtain
\begin{eqnarray}
q_{d|c}^{\ess} = \frac{\lambda_d^{\ess}w_{d|c}^{1/\theta}}{ \sum_a \lambda_{a}^{\ess}w_{a|c}^{1/\theta}} \ .
\label{sieq:qesslambda_ono}
\end{eqnarray}
This expression has a transparent interpretation. The evolutionary stable pre-allocation values depend on the transition probability $\bf{w}$ and linearly on the growth rate under that condition. The dependency on the transition probability is influenced by the shape of the trade-off. In the case $\theta=1$ $and$ if the growth rates are identical, cells allocate a fraction of the proteome exactly equal to the associated probability.
\begin{equation}
q_{d|c}^{\ess} = w_{d|c}.
\end{equation}

The values of $\lambda_a^{\ess}$ are determined by the solutions of
equation~\ref{sieq.lambdaess}, which reads
\begin{eqnarray}
   0 && =
\tau\sum_{a,b} \lambda_a w_{a|b} p_b \left(  
 \frac{\delta_{b,c}}{\lambda^*-\lambda_b} + \frac{\delta_{a,c}}{\lambda_a}
\right) \left( q_{a|b} \frac{\lambda^*-\lambda_b}{\lambda_a} \frac{\lambda_{\max}}{\lambda_{\max} - \lambda^*}\right)^{1-\theta} -\sum_a \lambda_a p_a \ \frac{\log  \bar{D}_c}{\lambda_c^2} \delta_{a,c} = \\ \nonumber
&& = 
\tau \sum_a \frac{\lambda_a w_{a|c} p_c}{\lambda^*-\lambda_c} \left( q_{a|c} \frac{\lambda^*-\lambda_c}{\lambda_a} \frac{\lambda_{\max}}{\lambda_{\max} - \lambda^*}\right)^{1-\theta} +
\tau \sum_{b} \frac{\lambda_c w_{c|b} p_b}{\lambda_c}
 \left( q_{c|b} \frac{\lambda^*-\lambda_b}{\lambda_c} \frac{\lambda_{\max}}{\lambda_{\max} - \lambda^*}\right)^{1-\theta} \\ \nonumber
&& - \frac{p_c \log  \bar{D}_c}{\lambda_c}
\ .
\label{sieq.zerolambda}
\end{eqnarray}
Under symmetric conditions ($w_{a|b}=1/{M-1}$ and $\log \bar{D}_a = \mathbb{E}(\log D)$) this equation is solved by eq.~\ref{sieq.lambdaessstochsym}, which corresponds to the minimization of the expected depletion time.

The solution of this equation becomes particularly simple when $\theta=1$. In that case, it reduces to
\begin{eqnarray}
   0 =
   \frac{\tau p_c}{\lambda^*-\lambda_c} \sum_a \lambda_a w_{a|c}    +
\tau \sum_{b} w_{c|b}p_b
- \frac{p_c \log  \bar{D}_c}{\lambda_c}
\ ,
   \\
\end{eqnarray}
from which one obtains
\begin{eqnarray}
   0 && =
\frac{1}{\lambda^*-\lambda_c} \sum_{a} \lambda_a w_{a|c} + 1
- \frac{ \log  \bar{D}_c}{\tau \lambda_c}  = \\
&& =
\frac{\lambda_c }{\lambda^*-\lambda_c}   +
\frac{1}{\lambda^*-\lambda_c}\sum_a(\lambda_a-\lambda_c)w_{a|c} + 1 -\frac{\log \bar D_c}{\tau \lambda_c}
\ ,
   \\
\end{eqnarray}
which is (implicitly) solved by
\begin{equation}
  \frac{\lambda^*-\lambda_c^{\ess}}{\lambda^{\ess}_c} = \frac{\tau}{\log \bar D_c} \left(\lambda^* + \sum_a (\lambda_a^{\ess}-\lambda_c^{\ess})w_{a|c}\right).
\end{equation}

\begin{equation}
  \frac{\lambda^*}{\lambda^{\ess}_c}  = 1 + \frac{\lambda^* \tau}{\log \bar D_c} \left(1+ \sum_a \frac{\lambda_a^{\ess}-\lambda_c^{\ess}}{\lambda^*}w_{a|c}\right) =
   1 + \frac{\lambda^* \tau}{\log \bar D_c} \left(1+ \frac{\mathbb{E}_{\text{next}|c}\left(\lambda^{\ess}\right)-\lambda_c^{\ess}}{\lambda^*}\right)
  \ ,
\end{equation}
where $\mathbb{E}_{\text{next}|c}\left(\lambda^{\ess}\right) = \sum_a \lambda_a w_{a|c}$ is the expected growth rate in the condition following $c$.

If $\lambda^{\ess}_a = \lambda^{\ess}_c = \lambda^{\ess}$ this equation reduces to the solution obtained in eq.~\ref{eq.lambdaessfluct}. This for instance occurs if $\bar{D}_a$ is constant, independent of condition. If that is not the case, i.e. if $\tilde{D}_a$ is condition dependent, then value of $\lambda^{\ess}_c$ is also influenced by the transition probability $\mathbf{w}$ trough the term $\sum_a (\lambda_a - \lambda_c) w_{a|c}$. This last term (which depends on $\lambda_c$ itself, therefore making the equation implicit) can be interpreted as the expected difference in growth rate between the current condition $c$ and the future (unknown) condition.
In particular, the value of $\lambda_c^{\ess}$ is smaller whenever the expected growth rate change is larger.
The value of $\lambda$ depends on the quality of the condition (value of $\log D$) and on the expected value of the growth in the new condition $\mathbb{E}_{\text{next}|c}\left(\lambda^{\ess}\right)$.
If all the values of $\lambda$ are equal to each other, one obtains the solution
\begin{eqnarray}
   \frac{\lambda^*}{\lambda^{\ess}} = 1 + \frac{\tau \lambda^*}{\mathbb{E}[\log D]} \ ,
\end{eqnarray}
which is the one reported in Fig.~\ref{fig:main2} and ~\ref{fig:main3}.

\subsection{Dynamics of pre-allocations for $\theta = 1$ and symmetric conditions}
\label{sisec.learning}

In the case $\theta = 1$, the lag time depends on pre-allocation through the function $G(x)=-\log x$. Using eq~\ref{sieq.expecteddepl_condexpl} we can write
\begin{equation}
 \frac{\partial}{\partial q_{c|d}}\mathbb{E}_a \left[ t_{s}\left( \underline{\lambda}^{\ess} , \mathbf{q}^{\ess} \right) \right] = 
- \frac{\partial}{\partial q_{c|d}} \frac{\tau}{p_a} \sum_b w_{a|b} p_b \log q_{a|b} \ ,
\end{equation}
and then, using eq.~\ref{eqsi.adaptiveq} we obtain
\begin{eqnarray}
  \frac{\partial \langle q_{d|c} \rangle}{\partial t}  \propto 
   \frac{\partial}{\partial q_{c|d}}  \sum_{a,b} w_{a|b} p_b \log q_{a|b} - Z_c =
   \frac{\partial}{\partial q_{c|d}}
   \sum_b p_b \left(
   \sum_{a} w_{a|b} \log \frac{q_{a|b}}{w_{a|b}}
   \right) - Z_c
   \ .
   \label{sieq:KLderivation}
\end{eqnarray}
We can define the Kullback-Leibler divergence between $w_{\cdot|b}$ and $q_{\cdot|b}$ as
\begin{eqnarray}
  D^{KL}_{b} := \sum_a w_{a|b} \log \frac{w_{a|b}}{q_{a|b}}
   \ ,
\end{eqnarray}
and the KL divergence averaged over the conditions
\begin{eqnarray}
  D^{KL}(\mathbf{q}|\mathbf{w}) :=
  \sum_b p_b D^{KL}_{b}
  \ .
\end{eqnarray}
by inserting this definition in eq.~\ref{sieq:KLderivation}, we obtain eq.~\ref{eq.KLdiv}.

Similarly, eq.~\ref{eq.Tlafmin} can be obtained using the fact that, for $G(x)=-\log x$, the expected lag-time under condition $a$ depends on $\mathbf{q}$ as
\begin{equation}
\mathbb{E}_a \left[ T\left( \underline{\lambda} , \mathbf{q} \right) \right] = \sum_b w_{b|a} T_{b|a}\left( \underline{\lambda} , \mathbf{q} \right) = \sum_b w_{b|a} \log q_{b|a} + c \ ,
\end{equation}
where $c$ depends on the $\underline{\lambda}$ and other parameters.
Substituting the evolutionary stable solution $\mathbf{q}^\ess = \mathbf{w}$, one obtains
\begin{equation}
\mathbb{E}_a \left[ T\left( \underline{\lambda}^\ess , \mathbf{q}^\ess \right) \right]  = -\sum_b w_{b|a} \log w_{b|a} + c \ ,
\end{equation}
where $c$ is independent of $\mathbf{w}$. By averaging over conditions, one obtains
\begin{equation}
\mathbb{E} \left[ T\left( \underline{\lambda}^\ess , \mathbf{q}^\ess \right) \right]  =
\sum_a p_a \mathbb{E}_a \left[ T\left( \underline{\lambda}^\ess , \mathbf{q}^\ess \right) \right]  = -\sum_a p_a \sum_b w_{b|a} \log w_{b|a} + c \ .
\end{equation}

\section{Numerical simulations}
\label{sisec.sims}

\subsection{Physiology and tradeoff}

\begin{figure}[tbp]
	\centering
	\includegraphics[width=\textwidth]{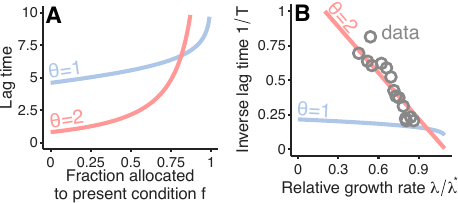} 
	\caption{ 
    Panel A) plots the lag-time as function of the allocation value $f$ for different tradeoff shapes: $\theta=1$ (blue, same as Fig~\ref{fig:main1}C) and  $\theta=2$ (red).
    Panel B) shows the tradeoff between inverse lag-time and growth rate for $\theta=1$ (blue, same as Fig~\ref{fig:main1}D) and  $\theta=2$ (red). Gray points are data from ref.~\cite{basan2020universal} for the lag-time between different glycolitic carbon sources to acetate.
    For both panels $\lambda_{\max} = 2.85 h^{-1}$, $\lambda^* = 1.1 h^{-1}$, $q=0.01$. In the case $\theta=1$ the lag timescale is $\tau=1 h$ and for $\theta=2$, $\tau=0.008 h$.
    }
\label{sfig:fig1boththeta}
\end{figure}

The reference parameters for the lag-growth tradeoff (Fig.~\ref{fig:main1} and~\ref{sfig:fig1boththeta}) are chosen mainly based on the scenario studied in~\cite{basan2020universal} (which corresponds to the shape $\theta=2$).

We consider $\lambda_{\max}=2.85 h^{-1}$ (based on estimates of $\phi_{\max}$ and translational capacity $\kappa_t$~\cite{scott2010interdependence}) and $\lambda^{*}=1.1 h^{-1}$ (which correspond to the critical value of the growth rate $\lambda$ at which the inverse lag-time tends to zero~\cite{basan2020universal}. Only the ratio $\lambda^*/\lambda_{\max}$ affects the evolutionary outcomes by shaping the dependency of lag-time on allocation values~\ref{sfig:lambdaTvsf}.

\subsection{Ecological dynamics within cycles}

Within a cycle, a strain $i$ starts at abundance $n^d_i(0)$. Its allocations $\underline{f}_i$ and $\mathbf{q}_i$ (together with the identity of the condition in cycle $d$ and $d-1$) determine the lag-time $T^d_i$ and growth rate $\lambda^d_i$.
We calculate the depletion time by solving numerically the equation
\begin{equation}
 c_0^d = \sum_i n_i^d(0) e^{\lambda^d_i(t-T^d_i)} \ , 
 \label{eq:depletiontime}
\end{equation}
where $c^d_0$ are the supplied resources.

We then calculate the initial abundance $n^{d+1}_i(0)$ of the following cycle as
\begin{equation}
n^{d+1}_i(0) = n^d_i(0) e^{\lambda_i(t-T^d_i)} / D^d \ , 
 \label{eq:depletiontime}
\end{equation}
where $D^d$ is the dilution factor, and $t$ is the solution of eq.~\ref{eq:depletiontime}.

Unless specified otherwise, the standard parameters are $D = 10^3$ and $c_0 = 1$. When $D$ and $c_0$ are randomly distributed we consider lognormal distribution with $\mathbb{E}(c_0)=1$ and $\exp(\mathbb{E}(\log D))=10^3$. We consider a stran extinct when $n_i(0) \leq 10^{-7}$.

\subsection{Environmental transitions}
\label{secsi.matrixw}

We generate the environmental transition matrices $\mathbf{w}$ from a Dirichlet distribution with parameter $\alpha$.
The probability distribution of the transition probability from condition $a$ to another condition is given by
\begin{equation}
P( \{w_{\cdot|a}\} ) \propto \prod_{b \neq a} (w_{b|a})^{\alpha} \ ,
\end{equation}
where the proportionality constant given by normalization is a Beta function and we implement the constraint $\sum_{b \neq a} w_{b|a}=1$. The probabilities are drawn independently across the starting condition $a$. The case $\alpha \to \infty$ correspond to the uniform case $w_{b|a}=1/(M-1)$ (for $b \neq a$). When $\alpha = 1$, the values $w_{\cdot|a}$ are uniformly distributed on a simplex. For values $\alpha > 1$ they tend to be more and more concentrated around $1/(M-1)$, while for $\alpha < 1$ one value becomes dominant respect to the others.

\subsection{Evolutionary dynamics}

We introduce one mutant at the beginning of a cycle with some probability $p_U$. We consider the number of births that occurred in the previous cycle $n^{d-1}_i(t_d)-n^{d-1}_i(0)$ from each strain and select a strain $i^*$ proportionally to this quantity. A mutant is introduced at a (low) abundance $n_{\text{inv}}$ and its parameters are chosen around the ones of the mother strain $\underline{f}_{i^*}$ and $\mathbf{q}_{i^*}$. In particular, we consider $\tilde{f}_{a,\text{mut}}=f_{a,i^{*}} + \sigma_{\text{mut}} \xi_a$ and $\tilde{q}_{a|b,\text{mut}}=q_{a|b,i^{*}} + \sigma_{\text{mut}} \zeta_{a|b}$, where $\xi$s and $\zeta$s are independent Gaussian random variables with mean zero and variance one. We obtain the allocations of the mutants $\underline{f}_{\text{mut}}$ and $\mathbf{q}_{\text{mut}}$ from $\tilde{\underline{f}}_{\text{mut}}$ and $\tilde{\mathbf{q}}_{\text{mut}}$ by imposing the normalization constraints and the boundary conditions (e.g., setting $f$ to 0 if $\tilde{f}$ becomes negative or to $1$ if $\tilde{f}>1$).

Unless specified otherwise, we consider $\sigma_{\text{mut}} = 0.05$, $p_U=1/2$ and $n_{\text{inv}}=10^{-5}$.

\newpage
\clearpage

\begin{figure}[tbp]
	\centering	\includegraphics[width=\textwidth]{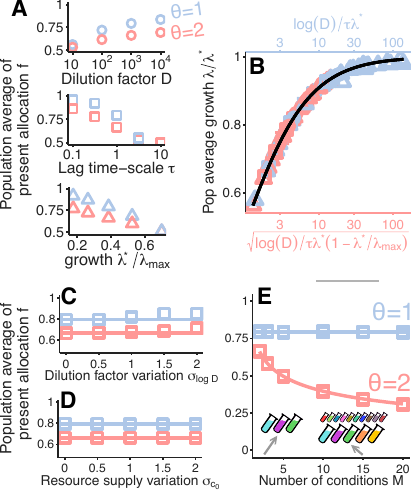}
	\caption{
The panels are obtained in the same way as Fig.~\ref{fig:main2} and Fig.~\ref{fig:main3} for $\theta=1$ (blue, same in the main text) and $\theta=2$ (red). The population average growth rate of evolved populations (colored points) depends on the ratio of lag and growth phase timescales for both values of $\theta$ (panel B). The functional dependency is different and matched was predicted obtained in eq.~\ref{si.eq.lambdaess}.
Panel C, D, E describe the effect of stochasticity on the outcome of evolutionary dynamics (corresponding to Fig.~\ref{fig:main3}). The variability of the logarithm of dilution factor ($\sigma_{\log D}$) and the concentration ($\sigma_{c_{0}}$) have no effect on the final allocation value $f$ for both $\theta=1$ and $\theta = 2$. Panel E) shows that the number of conditions $M$ affect the 
evolution of $f$ (numerical simulations, points) for $\theta \neq 1$ as predicted by eq.~\ref{sieq.lambdaessstochsym} (colored lines).
}
\label{sfig:fig2boththeta}
\end{figure}

\begin{figure}[tbp]	\centering
	\includegraphics[width=\textwidth]{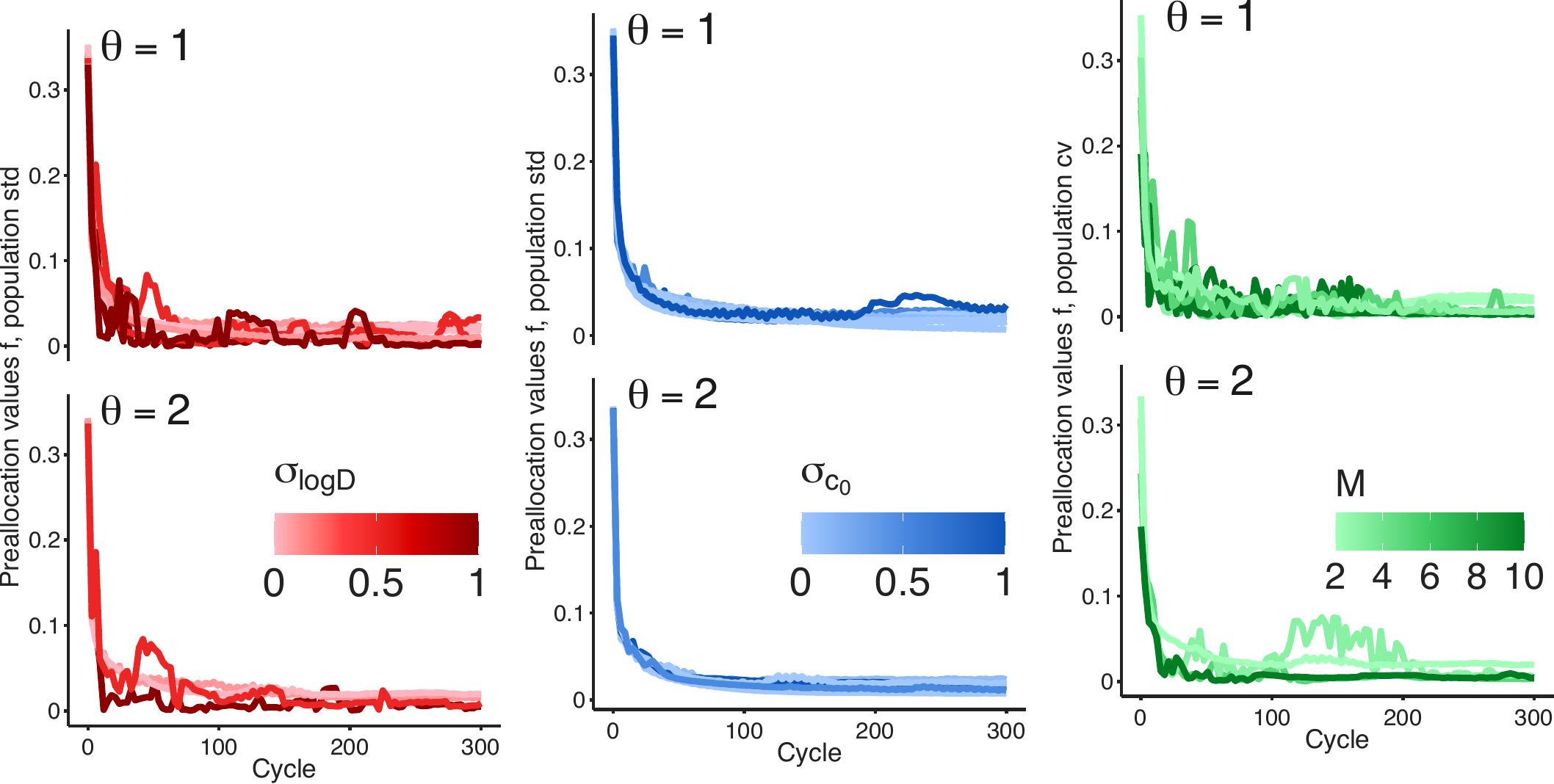} 
	\caption{
Each strain in the population is characterized by a typical allocation value $\bar{f}=\sum_a f_a/M$. The panels report the population variance of this quantity over time for different level of stochasticity ans tradeoff shapes. If strain $i$ has allocation $\bar{f}_i$ and relative abundance $x_i$, the population variance is defined as $\sum_i x_i (\bar{f}_i)^2 - (\sum_i x_i \bar{f}_i)^2$. }
\label{sfig:varianceevolving1}
\end{figure}

\begin{figure}[tbp]
	\centering
	\includegraphics[width=\textwidth]{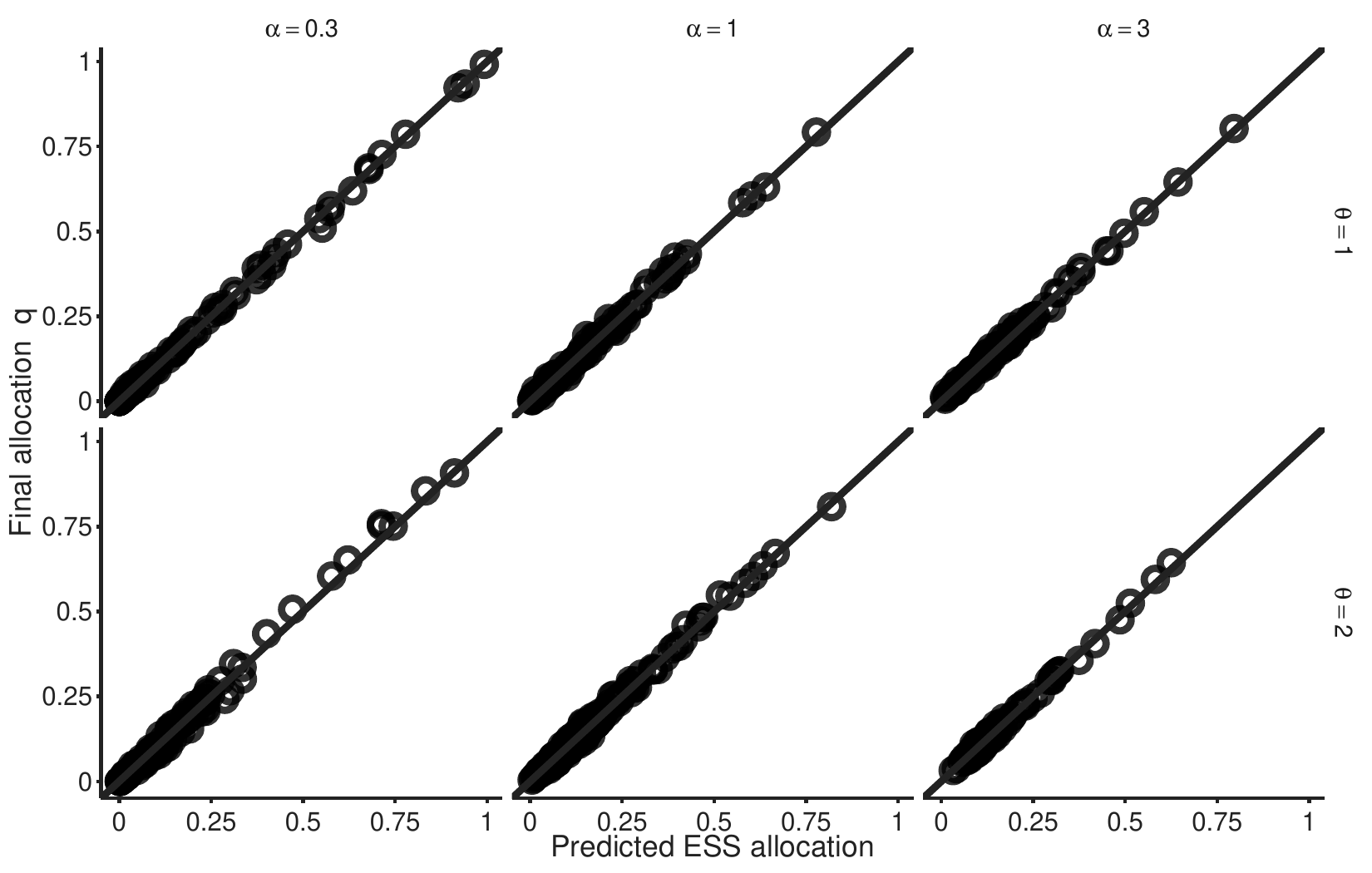} 
	\caption{
The panels compare the evolved pre-allocation values $q_{a|b}$ with the predicted ones obtained using equation~\ref{sieq:qesslambda_ono} assuming $\lambda^{\ess}_a$ constant indpendent of $a$. In this case the prediction reduces to $q_{a|c}^{\ess} \propto (w_{a|c})^{1/\theta}$. Different rows corresponds to different tradeoff shapes (parameterized by $\theta$) and different levels of environmental uncertainty (columns, parameterized by $\alpha$, as defined in section~\ref{secsi.matrixw}). While the approximation $q_{a|c}^{\ess} \propto (w_{a|c})^{1/\theta}$ is correct only for $\theta=1$ or $\alpha \to \infty$ (constant case, rightmost column, corresponding to a completely unstructured environment), it approximates well also the case $\theta = 2$ with low values of $\alpha$.
    }
\label{sfig:fig4theta}
\end{figure}

\begin{figure}[tbp]
	\centering
	\includegraphics[width=\textwidth]{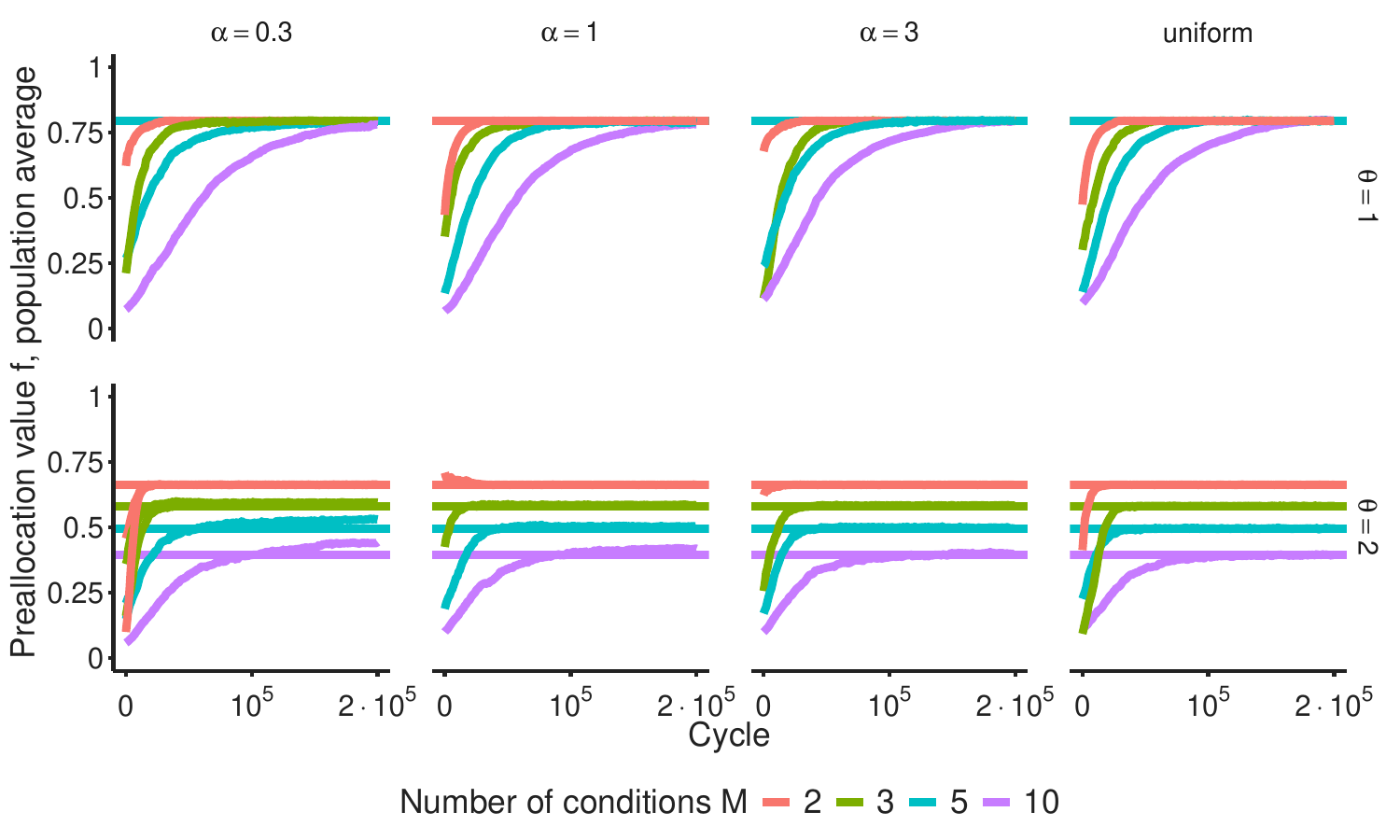} 
	\caption{
Each strain in the evolving population is characterized by a typical allocation value $\bar{f}=\sum_a f_a/M$. The panels report the population average of this quantity.
If strain $i$ has allocation $\bar{f}_i$ and relative abundance $x_i$, the population average is defined as $\langle \bar{f} \rangle = \sum_i x_i \bar{f}_i$. Different panels correspond to
different shapes of the tradeoff (rows, parameterized by $\theta$) and levels of environmental uncertainty (columns, parameterized by $\alpha$, as defined in section~\ref{secsi.matrixw}). The horizontal lines are the predictions obtained assuming symmetric environment and minimization of the depletion time (i.e., corresponding to~\ref{sieq.lambdaessstochsym}). The symmetric solution is exact for $\theta = 1$ or for constant transition matrix ($\alpha \to \infty$, rightmost column).
    }
\label{sfig:mean_f_time}
\end{figure}

\begin{figure}[tbp]	\centering
	\includegraphics[width=\textwidth]{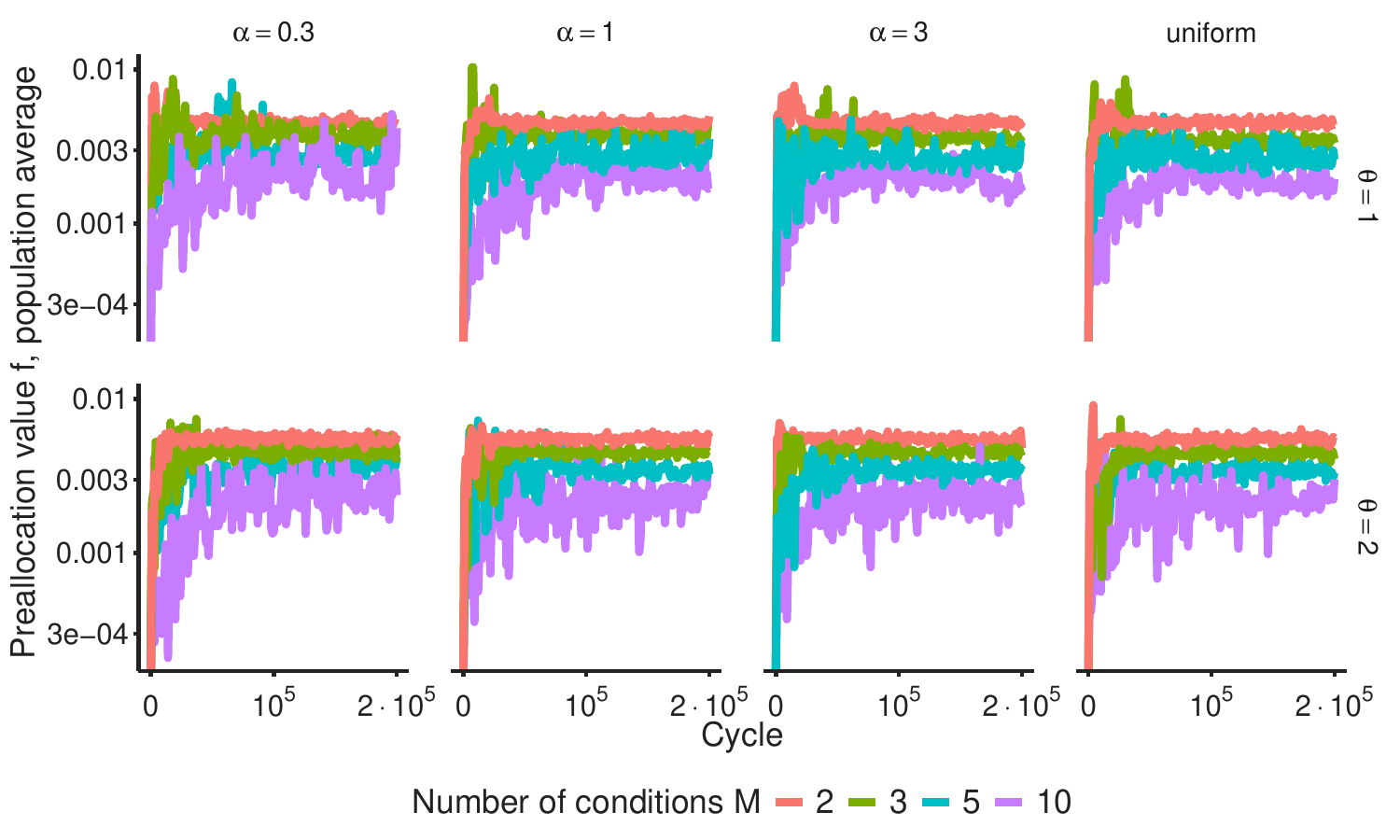} 
	\caption{
Each strain in the population is characterized by a typical allocation value $\bar{f}=\sum_a f_a/M$. The panels report the population variance of this quantity. If strain $i$ has allocation $\bar{f}_i$ and relative abundance $x_i$, the population variance is defined as $\sum_i x_i (\bar{f}_i)^2 - (\sum_i x_i \bar{f}_i)^2$. Across shapes of the tradeoff (rows, parameterized by $\theta$) and levels of environmental uncertainty (columns, parameterized by $\alpha$, as defined in section~\ref{secsi.matrixw}) the population variance stays to low values indicating a monomorphic population with no coexistence of multiple strains.}
\label{sfig:varianceevolving2}
\end{figure}

\begin{figure}[tbp]
	\centering
	\includegraphics[width=\textwidth]{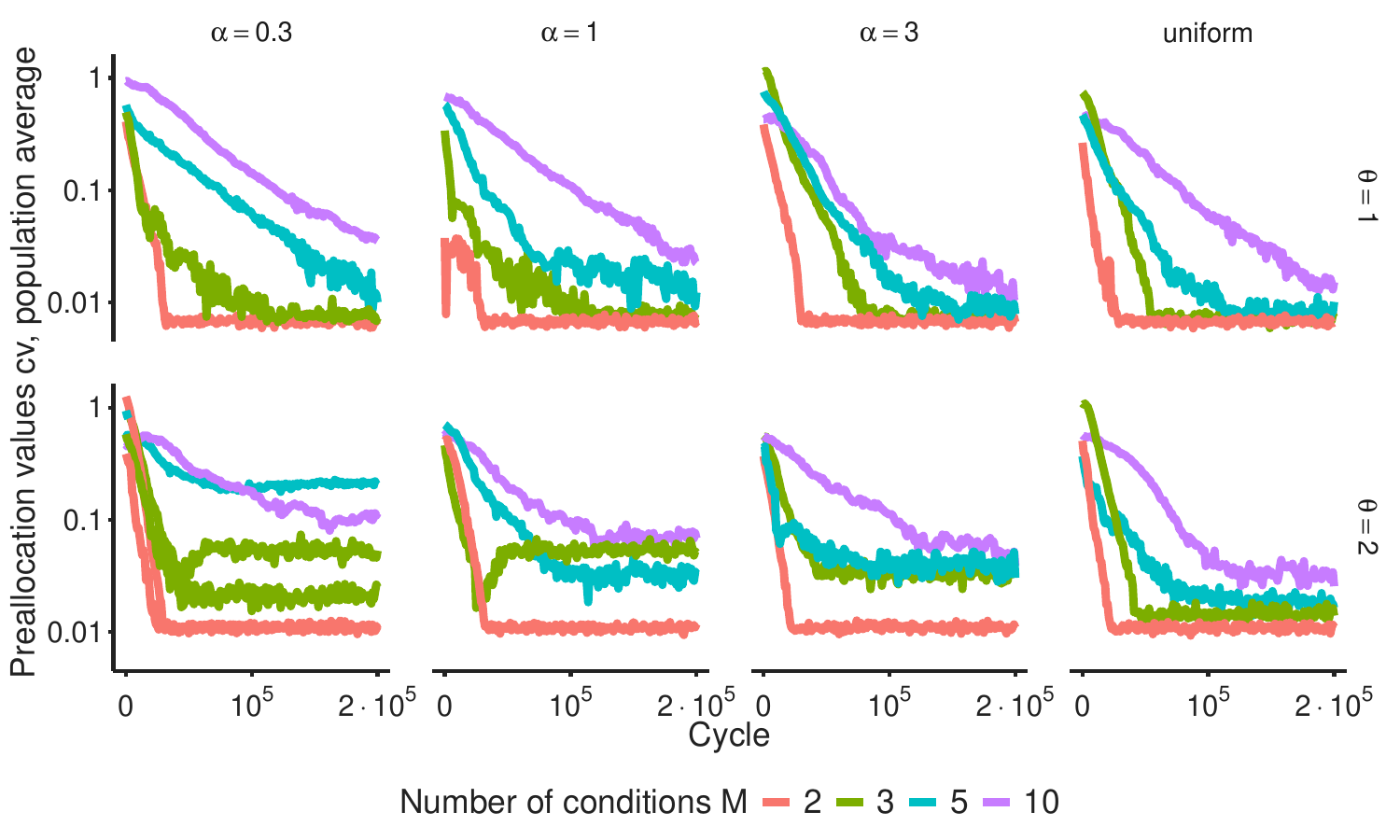} 
	\caption{
Each strain in the population is characterized by an allocation value $f_a$ in condition $a$. These allocations are in general different from different conditions. We can quantify this variability with the coefficient of variation $\text{cv}(\underline{f}) =M \sum_a f_a^2 / (\sum_a f_a)^2 - 1$.  If strain $i$ has allocation $\underline{f}_i$ and relative abundance $x_i$, the population variance of the coefficient of variation is defined as $\langle \text{cv}(\underline{f}_i) \rangle =  \sum_i x_i  \text{cv}(\underline{f}_i)$. Different panels reports the value of $\langle \text{cv}(\underline{f}_i) \rangle$ over evolutionary times for different shapes of the tradeoff (rows, parameterized by $\theta$) and levels of environmental uncertainty (columns, parameterized by $\alpha$, as defined in section~\ref{secsi.matrixw}). A values $\langle \text{cv}(\underline{f}_i) \rangle$ corresponds to fully symmetric solution $f_a = f$ for all conditions $a=1,\dots,M$. As expected this is achieved for $M=2$, for $\theta = 1$ or for constant transition matrix ($\alpha \to \infty$, rightmost column).
    }
\label{sfig:sd_f_time}
\end{figure}

\clearpage

\end{appendix}

\end{document}